\documentclass[useAMS,usenatbib]{mn2e}

\usepackage{graphicx}
\usepackage{epsfig}  
\usepackage{times}    
\usepackage{subfigure}
\title[A New Look at Massive Clusters]{A New Look at Massive Clusters: weak lensing constraints on the triaxial dark matter halos of A1689, A1835, \& A2204}
\author[V. L. Corless, L. J. King, \& D. Clowe, ]{Virginia L. Corless,$^{1}$\thanks{E-mail:
vc258@ast.cam.ac.uk} Lindsay J. King$^{1}$, \& Douglas Clowe$^{2}$\\
$^{1}$Institute of Astronomy, University of Cambridge, Madingley Road, Cambridge, United Kingdom\\
$^{2}$Department of Physics and Astronomy, Ohio University, Athens, Ohio, USA}

\begin{document}

\date{2008 May 11}

\pagerange{\pageref{firstpage}--\pageref{lastpage}} \pubyear{2008}

\maketitle

\label{firstpage}

\begin{abstract}
Measuring the 3D distribution of mass on galaxy cluster scales is a crucial test of the $\Lambda$CDM model, providing constraints on the nature of dark matter.  Recent work investigating mass distributions of individual galaxy clusters (e.g. Abell 1689) using weak and strong gravitational lensing has revealed potential inconsistencies between the predictions of structure formation models relating halo mass to concentration and those relationships as measured in massive clusters.  However, such analyses employ simple spherical halo models while a growing body of work indicates that triaxial 3D halo structure is both common and important in parameter estimates.  We recently introduced a Markov Chain Monte Carlo (MCMC) method to fit fully triaxial models to weak lensing data that gives parameter and error estimates that fully incorporate the true shape uncertainty present in nature. In this paper we apply that method to weak lensing data obtained with the ESO/MPG Wide-Field Imager for galaxy clusters A1689, A1835, and A2204, under a range of Bayesian priors derived from theory and from independent X-ray and strong lensing observations. For Abell 1689, using a simple strong lensing prior we find marginalized mean parameter values $M_{200} = (0.83 \pm 0.16)\times10^{15}$ $h^{-1}$M$_{\odot}$ and $C=12.2\pm 6.7$, which are marginally consistent with the mass-concentration relation predicted in $\Lambda CDM$.  With the same strong lensing prior we find for Abell 1835 $M_{200} =(0.67 \pm 0.22)\times10^{15}$ $h^{-1}$M$_{\odot}$ and $C=7.1^{+ 10.6}_{-7.1}$, and using weak lensing information alone find for Abell 2204 $M_{200} =(0.50 \pm 0.19)\times10^{15}$ $h^{-1}$M$_{\odot}$ and $C=7.1\pm 6.2$.  The large error contours that accompany our triaxial parameter estimates more accurately represent the true extent of our limited knowledge of the structure of galaxy cluster lenses, and make clear the importance of combining many constraints from other theoretical, lensing (strong, flexion), or other observational (X-ray, SZ, dynamical) data to confidently measure cluster mass profiles.  If we assume CDM to be correct and apply a mass-concentration prior derived from CDM structure formation simulations, we find \{$M_{200} =(0.99 \pm 0.18)\times10^{15}$ $h^{-1}$M$_{\odot}$; $C=7.7\pm 2.1$\}, \{$M_{200} =(0.68 \pm 0.19)\times10^{15}$ $h^{-1}$M$_{\odot}$; $C=4.4\pm 1.6$\}, and \{$M_{200} =(0.45 \pm 0.13)\times10^{15}$ $h^{-1}$M$_{\odot}$; $C=5.0\pm 1.7$\} for A1689, A1835, and A2204 respectively.
\end{abstract}

\section{Introduction}
\begin{figure*}
\centering
\rotatebox{0}{
\resizebox{16cm}{!}
{\includegraphics{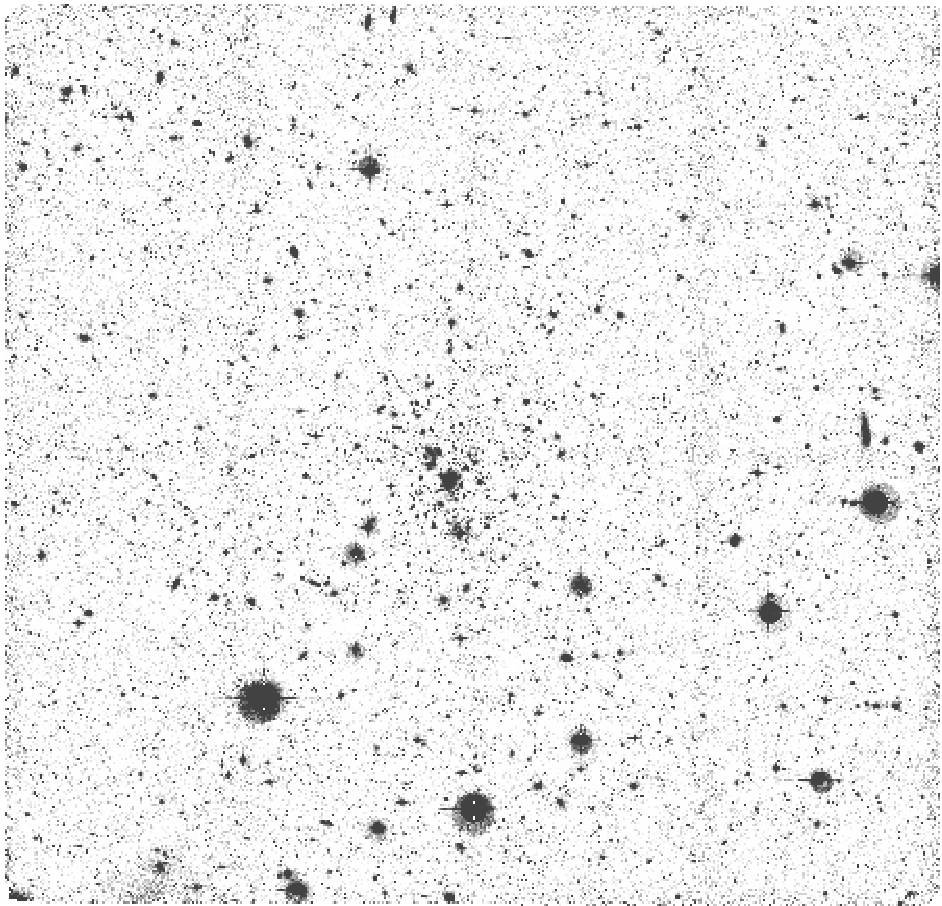}\includegraphics{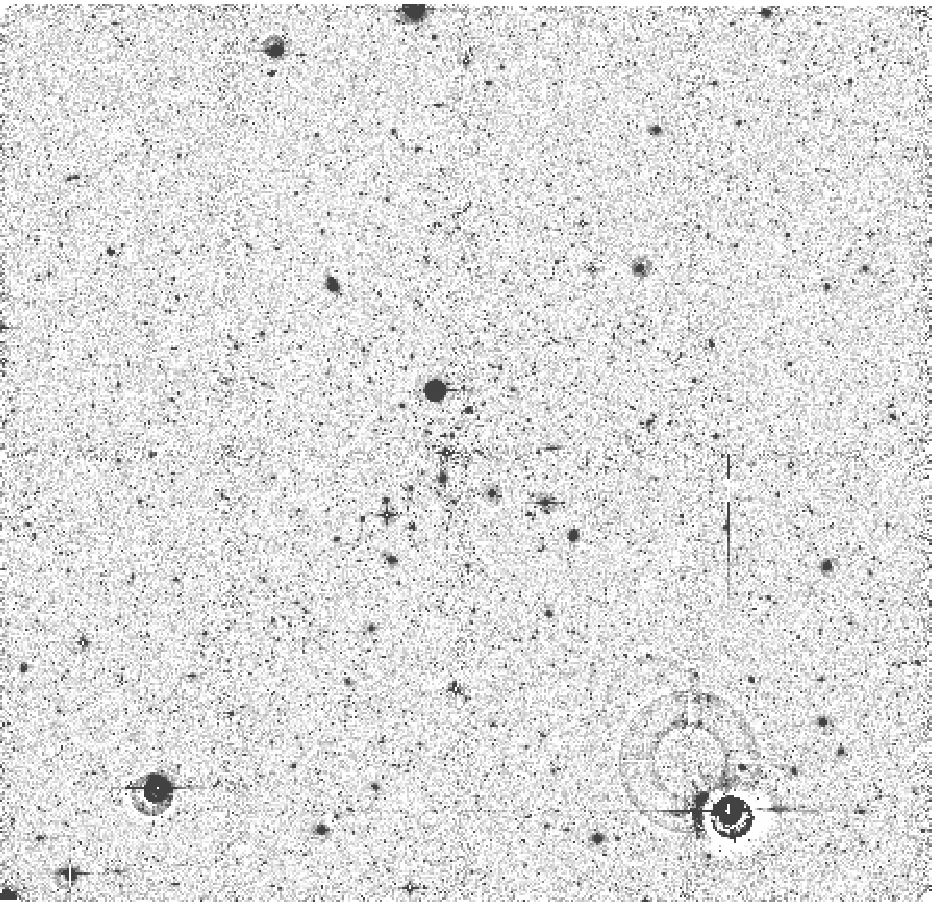}\includegraphics{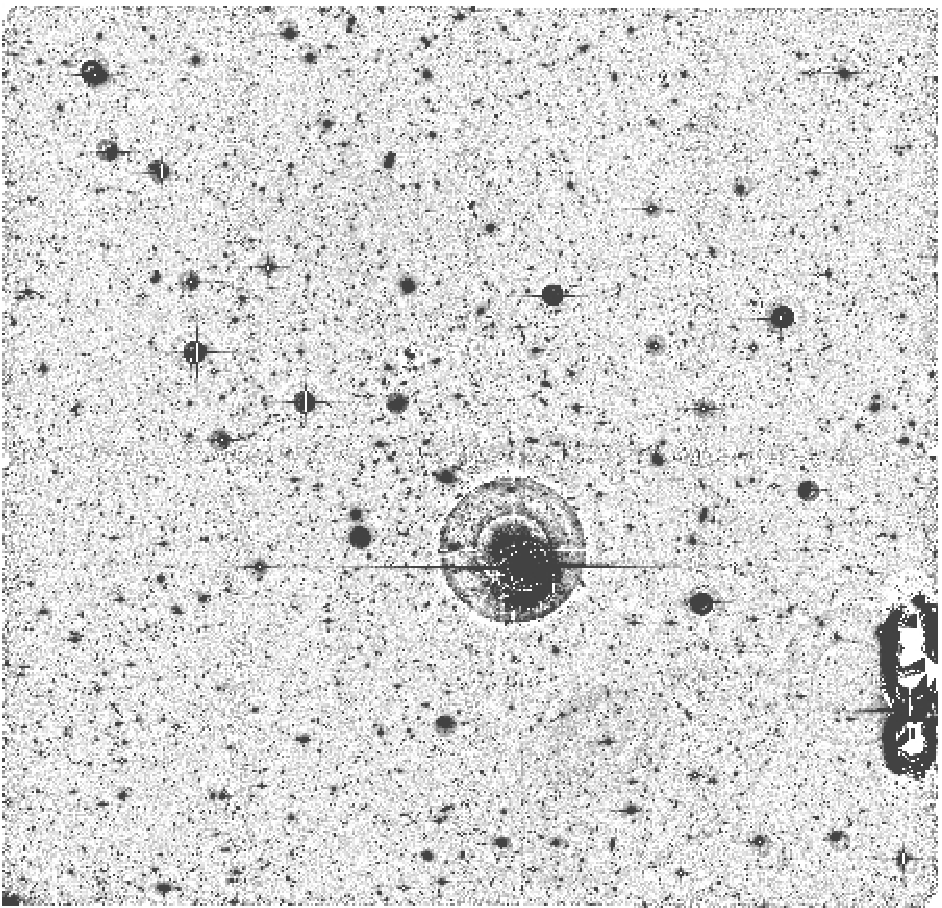}}
}
 \caption{34' $\times$ 34' $R$-band wide-field images of Abell 1689, 1835, and 2204, from the ESO/MPG WFI at La Silla.}
 \label{fig:plot1}
\end{figure*}
Galaxy clusters are ideal laboratories in which to study dark matter, being the most massive bound structures in the universe and dominated by their dark matter component ($\sim90\%$). Constraining the clustering properties of dark matter is crucial for refining structure formation models that predict both the shapes of dark matter halos and their mass function (e.g. \cite{navarro97}; \cite{evrard02}; \cite{bahcall03}; \cite{dahle06}).  Several methods are used to measure galaxy cluster dark matter profile shapes and halo masses on a range of scales, including X-ray and Sunyaev-Zeldovich (SZ) studies, dynamical analyses, and gravitational lensing.  However, all of these methods typically require simplifying assumptions to be made regarding the shape and/or dynamical state of the cluster in order to derive meaningful constraints from available data.  Crucially, while most parametric methods typically assume spherical symmetry of the halo, observed galaxy clusters often exhibit significant projected ellipticity and halos in CDM structure formation simulations (e.g. \cite{bett07} (using the Millennium simulation); \cite{shaw06}) show significant triaxiality in cluster-scale halos, with axis ratios between minor and major axes as small as 0.3.  Understanding and accurately incorporating the impact of this on cluster mass and parameter estimates is crucial for accurate comparisons between measured cluster properties and model predictions (e.g. the $\Lambda$CDM mass-concentration relation).

Efforts to understand the impact of triaxiality in gravitational lensing and its potential role in explaining apparent discrepancies with CDM began when \cite{oguri05} applied a fully triaxial NFW model to the shear map of Abell 1689 to find that it is consistent with $6\%$ of cluster-scale halos. These continued as \cite{gavazzi05} showed that a triaxial NFW can reconcile parameter values derived from observations of the cluster MS2137-23 to predictions from N-body simulations.

More generally, \cite{corless07} demonstrated in the weak lensing regime that neglecting halo triaxiality in parameterised fits of NFW models to dark matter halos with axis ratios significantly less than one can lead to over- and underestimates of up to 50$\%$ and a factor of 2 in halo mass and concentration, respectively.  While extreme cases of triaxiality are rare, such halos can be much more efficient lenses than their more spherical counterparts, especially when in configurations that hide most of the triaxial shape along the line of sight.  \cite{oguri08} recently quantified this expectation, showing that the strongest lenses in the universe are expected to be a highly biased population, strongly favouring orientations along the line of sight, high levels of triaxiality, and apparent sphericity in projection; these biases correspond to a population of very triaxial prolate (``cigar''-shaped) halos oriented with their long axes approximately aligned along the line-of-sight.  Thus, for strong lensing clusters -- even those that are circular in projection on the sky such as Abell 1689 -- spherical symmetry is an unjustified and error-inducing assumption.

Further, even halos with less extreme axis ratios are inaccurately fit by spherical models.  We expect the vast majority of galaxy cluster scale dark matter halos to be triaxial, and so it is important to include this expectation in model fitting.  Moreover, \cite{corless08} (herein CK08) showed that triaxiality leaves no quantifiable signature in lensing data (i.e. in the  maximum-likelihood values obtained fitting models), and so it is generally not possible to determine {\it a priori} the expected impact of triaxiality on mass and concentration estimates for a given lensing halo.

In the past, triaxial models have not generally been fit to weak lensing data because they cannot easily be well-constrained, largely due to the intrinsic limitations of lensing.  Because lensing is determined only by the projected mass density and shear of the underlying mass distribution, it is inherently impossible to fully constrain a 3D structure without imposing strong priors on the shape of the halo or supplementing lensing data with other data types more sensitive to line-of-sight halo structure, such as dynamical information.  

\begin{table}
\centering
\caption{Properties of the three lensing clusters}
\label{tab:table1}
\begin{tabular}[t!]{ccccccccccccc}
\hline
Cluster&RA&Dec&z&n [arcmin$^{-2}$]&$\sigma_{\epsilon}$\\
\hline
A1689&13 11 29.5&-01 20 17&0.1832&7&0.34\\
A1835&14 01 02.0&+02 51 32&0.2532&12&0.42\\
A2204&15 32 45.7&+05 34 43&0.1522&15&0.44\\
\hline
\end{tabular}
\end{table} 

Despite these difficulties, that triaxiality has been convincingly demonstrated to be an important factor in model parameter estimation in lensing analyses make it important to directly include it in NFW fits to galaxy clusters. For studies of individual clusters doing so is crucial if claims regarding the validity of the $\Lambda$CDM paradigm based on NFW parameter estimates are to be meaningfully evaluated.  It is also of importance across populations to obtain more accurate distributions of galaxy cluster parameters, and especially in measuring the galaxy cluster mass function and constraining the scatter in mass-observable relations.  If the mass function is significantly sloped as expected (e.g. \cite{evrard02}), even the best-case scenario of a symmetric scatter of mass estimates due to neglected triaxiality would lead to an asymmetric shift in the calculated mass function, because there are more low mass halos to shift up in mass than there are high mass halos to shift down (see e.g. \cite{corless08b}).

In CK08 we introduced a Bayesian Markov Chain Monte Carlo (MCMC) method for fitting triaxial NFW models to weak gravitational lensing data.  We demonstrated that it accurately recovers mean parameter values and errors across a population of triaxial dark matter halos similar to those seen in CDM structure formation simulations.  In this paper we make the first application of that triaxial fitting method to the weak lensing signals of three well-known galaxy clusters: Abell 1689, Abell 1835, and Abell 2204.  We reanalyze preexisting lensing data obtained with 34x34 arcmin observations of the three cluster fields with the ESO/MPG Wide Field Imager, first presented and modelled with spherical mass models in \cite{clowe01}, \cite{clowe02}, and \cite{king02a}.  

The MCMC method allows for the simple and statistically robust imposition of prior probability function on the parameters of the triaxial model.  Though some would rather avoid priors altogether, simple models include hidden priors stronger than those we employ in this work; for example, a ``simple'' spherical model includes implicit $\delta$-function priors on the triaxial axis ratios.  Priors are of such importance in this problem because models of 3D structures fit using only 2D lensing data are intrinsically underconstrained.  CK08 showed that carefully chosen priors are crucial to obtaining statistically accurate results across a population, but that slight offsets between the true distribution of parameters in nature and the distributions employed as priors lead only to a very small decline in the accurate performance of the triaxial fitting method.  We therefore continue in this paper the investigation of the behaviour and best use of various prior probability functions in fitting the weak lensing signals of galaxy clusters, in specific application to A1689, A1835, and A2204.

An additional advantage of the MCMC method is that it allows for the straightforward and statistically-robust inclusion of additional constraints from other independent and complementary observations such as strong lensing, SZ, X-ray and spectroscopic studies.  To demonstrate that capability, we investigate the inclusion of constraints from strong lensing and X-ray observations of the three clusters, imposing the external constraints as prior probability functions on the cluster model.

Section 2 briefly describes the observational data, reviews lensing by triaxial dark matter halos and the MCMC triaxial fitting method, and discusses the various priors employed in this paper.  Section 3 presents our results, and Section 4 concludes with some discussion. Throughout we assume a concordance cosmology with $\Omega_m=0.3$, $h=0.7$, and a cosmological constant $\Omega_{\Lambda} = 0.7$.

\begin{figure*}
\centering
\rotatebox{0}{
\resizebox{13cm}{!}
{\includegraphics{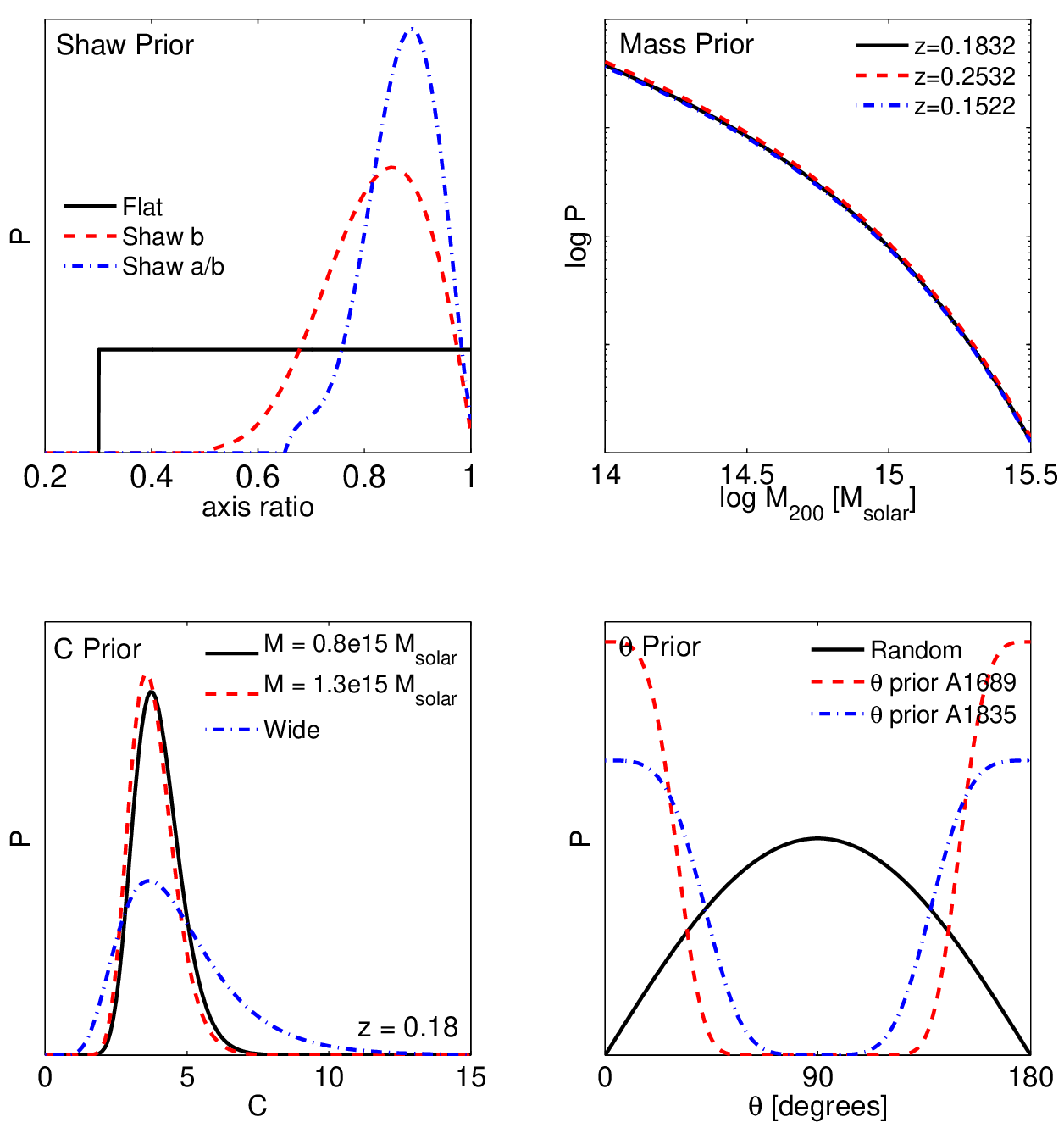}}
}
 \caption{Prior probability distributions of axis ratios (Shaw), virial mass (Mass), mass-concentration (C), and triaxial line-of-sight orientation angle ($\theta$), all derived from predictions of $\Lambda$CDM structure formation or lensing simulations.}
 \label{fig:plot2}
\end{figure*}

\begin{table*}
\centering
\caption{The marginalized mean parameter values obtained fitting single triaxial NFWs to A1689, A1835, and A2204 under a range of prior probability distributions on the halo mass, concentration, shape and orientation derived from CDM structure formation simulations and external observational constraints, described in Sec. \ref{sec:priors}.  $1\sigma$ errors on the estimate of the mean parameter values are given; all values are for a concordance cosmology of $\Omega_m = 0.3$, $\Omega_{\Lambda}=0.7$, $h=0.7$.}
\label{tab:table2}
\begin{tabular}[t!]{ccccccccccccc}
\hline
Prior&$M_{200} [10^{15}M_{\odot}]$&$C$&$a$&$b$&$M_{ES}$&$C_{ES}$\\
\hline
A1689&&&&\\
Flat&$1.31 \pm 0.28$&$15.4\pm 9.0$&$0.55\pm 0.19$&$0.82\pm0.14$&$1.28^{+0.30}_{-0.35}$&$15.3\pm9.1$\\
Shaw&$1.27 \pm 0.25$&$13.7\pm 7.7$&$0.74\pm 0.09$&$0.85\pm0.08$&$1.27^{+0.26}_{-0.26}$&$13.7\pm7.7$\\
Mass&$1.15 \pm 0.24$&$19.5\pm 12.3$&$0.56\pm 0.19$&$0.81\pm0.15$&$1.13^{+0.26}_{-0.30}$&$19.4^{+12.3}_{-12.4}$\\
C&$1.72 \pm 0.37$&$4.0\pm 0.5$&$0.53\pm 0.20$&$0.71\pm0.20$&$1.65^{+0.41}_{-0.48}$&$4.0^{+0.6}_{-0.7}$\\
MC&$1.41 \pm 0.25$&$7.7\pm 2.1$&$0.59\pm 0.19$&$0.78\pm0.17$&$1.38^{+0.27}_{-0.31}$&$7.7^{+2.1}_{-2.2}$\\
Sphere&$1.27\pm0.24$&$13.3\pm7.6$&&&&\\
$\theta$&$1.18 \pm 0.23$&$12.2\pm 6.7$&$0.55\pm 0.20$&$0.74\pm0.18$&$1.16^{+0.25}_{-0.30}$&$12.1^{+6.8}_{-6.8}$\\
$\theta_E$&$1.18 \pm 0.20$&$18.8\pm 6.5$&$0.55\pm 0.20$&$0.81\pm0.14$&$1.16^{+0.22}_{-0.25}$&$18.7^{+6.6}_{-6.7}$\\
$\theta + \theta_E$&$1.07 \pm 0.18$&$15.3\pm 5.2$&$0.55\pm 0.20$&$0.74\pm0.18$&$1.05^{+0.19}_{-0.22}$&$15.2^{+5.2}_{-5.3}$\\
$\theta_E$ + MC&$1.19 \pm 0.18$&$12.7\pm 2.6$&$0.54\pm 0.20$&$0.71\pm0.20$&$1.17^{+0.20}_{-0.24}$&$12.6^{+2.7}_{-2.8}$\\
\hline
A1835&&&&\\
Flat&$1.10 \pm 0.43$&$8.0^{+10.9}_{-8.0}$&$0.51\pm 0.19$&$0.77\pm0.16$&$1.07^{+0.45}_{-0.47}$&$7.9^{+11.0}_{-7.9}$\\
Shaw&$1.01 \pm 0.32$&$7.3^{+10.9}_{-7.3}$&$0.73\pm 0.09$&$0.84\pm0.09$&$1.01^{+0.32}_{-0.32}$&$7.2^{+10.9}_{-7.2}$\\
Mass&$0.83 \pm 0.28$&$13.3^{+17.6}_{-13.3}$&$0.52\pm 0.19$&$0.76\pm0.17$&$0.81^{+0.29}_{-0.30}$&$13.2^{+17.7}_{-13.2}$\\
C&$1.27 \pm 0.41$&$3.5\pm 0.6$&$0.53\pm 0.19$&$0.78\pm0.16$&$1.22^{+0.44}_{-0.47}$&$3.5^{+0.6}_{-0.7}$\\
MC&$0.97 \pm 0.27$&$4.4\pm 1.6$&$0.55\pm 0.18$&$0.78\pm0.16$&$0.94^{+0.29}_{-0.32}$&$4.4^{+1.6}_{-1.7}$\\
Sphere&$1.01\pm0.30$&$6.8^{+9.5}_{-6.8}$&&&&\\
$\theta$&$0.95 \pm 0.31$&$7.1^{+10.6}_{-7.1}$&$0.51\pm 0.20$&$0.74\pm0.18$&$0.92^{+0.34}_{-0.35}$&$7.0^{+10.6}_{-7.0}$\\
X-ray&$1.17 \pm 0.36$&$3.9\pm 1.0$&$0.54\pm 0.19$&$0.78\pm0.16$&$1.13^{+0.39}_{-0.43}$&$3.8^{+1.0}_{-1.1}$\\
X-ray $+\theta$&$0.98 \pm 0.25$&$3.7\pm 1.0$&$0.53\pm 0.20$&$0.74\pm0.18$&$0.95^{+0.28}_{-0.32}$&$3.7^{+1.0}_{-1.1}$\\
\hline
A2204&&&&\\
Flat&$0.72 \pm 0.27$&$7.1\pm 6.2$&$0.52\pm 0.20$&$0.81\pm0.15$&$0.70^{+0.29}_{-0.35}$&$7.1\pm6.2$\\
Shaw&$0.68 \pm 0.21$&$6.8\pm 5.8$&$0.74\pm 0.09$&$0.85\pm0.08$&$0.67^{+0.22}_{-0.22}$&$6.8\pm5.8$\\
Mass&$0.56 \pm 0.20$&$10.4\pm 9.7$&$0.53\pm 0.20$&$0.79\pm0.16$&$0.54^{+0.21}_{-0.26}$&$10.4^{+9.8}_{-9.7}$\\
C&$0.80 \pm 0.25$&$4.1\pm 0.8$&$0.54\pm 0.20$&$0.82\pm0.14$&$0.78^{+0.27}_{-0.29}$&$4.1^{+0.8}_{-0.9}$\\
MC&$0.64 \pm 0.19$&$5.0\pm 1.7$&$0.54\pm 0.20$&$0.81\pm0.15$&$0.62^{+0.20}_{-0.23}$&$5.0\pm1.8$\\
Sphere&$0.69\pm0.21$&$6.6^{+6.7}_{-6.6}$&&&&\\
\hline
\end{tabular}
\end{table*} 

\begin{table*}
\centering
\caption{The approximate maximum-likelihood parameter values obtained fitting single triaxial NFWs to A1689, A1835, and A2204 under a range of prior probability distributions on the halo mass, concentration, shape and orientation derived from CDM simulations and external observational constraints, described in Sec. \ref{sec:priors}.  $1\sigma$ (68\%) errors on the maximum-likelihood parameter values are given; all values are for a concordance cosmology of $\Omega_m = 0.3$, $\Omega_{\Lambda}=0.7$, $h=0.7$.}
\label{tab:table3}
\begin{tabular}[t!]{ccccccccccccc}
\hline
Prior&$M_{200} [10^{15}M_{\odot}]$&$C$&$a$&$b$&$M_{ES}$&$C_{ES}$&$\chi^2$&Reduced $\chi^2$\\
\hline
A1689&&&&\\
Flat&$1.52^{+0.96}_{-0.59}$&$12.4^{+24.2}_{-7.7}$&$0.31^{+0.68}_{-0.20}$&$1.00^{+0.00}_{-0.70}$&$1.38^{+1.09}_{-0.79}$&$12.0^{+24.6}_{-8.0}$&5953.59&1.04084\\
Shaw&$1.38^{+0.63}_{-0.48}$&$10.6^{+18.7}_{-5.3}$&$0.81^{+0.15}_{-0.29}$&$0.89^{+0.11}_{-0.25}$&$1.37^{+0.63}_{-0.52}$&$10.6^{+18.7}_{-5.4}$&5953.6&1.04084\\
Mass&$1.22^{+0.83}_{-0.39}$&$11.8^{+31.7}_{-6.7}$&$0.38^{+0.62}_{-0.26}$&$0.86^{+0.14}_{-0.56}$&$1.14^{+0.90}_{-0.60}$&$11.6^{+31.9}_{-7.1}$&5953.85&1.04088\\
C&$1.17^{+1.49}_{-0.32}$&$3.6^{+2.4}_{-0.9}$&$0.19^{+0.81}_{-0.09}$&$0.31^{+0.69}_{-0.01}$&$0.88^{+1.79}_{-0.42}$&$3.3^{+2.7}_{-1.1}$&5958.32&1.04167\\
MC&$1.11^{+1.01}_{-0.28}$&$6.0^{+6.0}_{-2.8}$&$0.16^{+0.84}_{-0.06}$&$0.34^{+0.66}_{-0.04}$&$0.85^{+1.27}_{-0.40}$&$5.5^{+6.5}_{-2.8}$&5955.05&1.04109\\
Sphere&$1.33^{+0.44}_{-0.34}$&$10.0^{+9.3}_{-4.0}$&&&&&5954.12&1.0402\\
$\theta$&$1.27^{+0.67}_{-0.55}$&$7.6^{+16.7}_{-4.4}$&$0.49^{+0.50}_{-0.39}$&$0.84^{+0.16}_{-0.54}$&$1.23^{+0.72}_{-0.83}$&$7.5^{+16.8}_{-4.9}$&5953.97&1.0409\\
$\theta_E$&$1.31^{+0.67}_{-0.44}$&$16.9^{+18.8}_{-8.5}$&$0.32^{+0.68}_{-0.20}$&$0.96^{+0.04}_{-0.64}$&$1.21^{+0.77}_{-0.59}$&$16.5^{+19.3}_{-8.9}$&5952.58&1.04066\\
$\theta + \theta_E$&$1.10^{+0.51}_{-0.37}$&$11.8^{+12.8}_{-5.9}$&$0.49^{+0.50}_{-0.39}$&$0.78^{+0.22}_{-0.48}$&$1.07^{+0.54}_{-0.61}$&$11.7^{+12.9}_{-6.6}$&5952.58&1.04066\\
$\theta_E$ + MC&$0.98^{+0.66}_{-0.21}$&$8.6^{+9.5}_{-1.5}$&$0.19^{+0.80}_{-0.09}$&$0.35^{+0.65}_{-0.05}$&$0.81^{+0.83}_{-0.29}$&$8.0^{+10.0}_{-1.9}$&5952.41&1.04063\\
\hline
A1835&&&&\\
Flat&$1.32^{+2.01}_{-0.73}$&$3.9^{+13.2}_{-2.8}$&$0.32^{+0.67}_{-0.22}$&$0.96^{+0.04}_{-0.65}$&$1.14^{+2.19}_{-0.91}$&$3.7^{+13.4}_{-2.9}$&8604.79&1.02134\\
Shaw&$1.19^{+0.70}_{-0.59}$&$3.6^{+10.8}_{-2.5}$&$0.73^{+0.23}_{-0.22}$&$0.83^{+0.17}_{-0.22}$&$1.18^{+0.72}_{-0.63}$&$3.5^{+10.8}_{-2.5}$&8605.71&1.02145\\
Mass&$0.86^{+1.05}_{-0.40}$&$4.0^{+22.0}_{-3.2}$&$0.50^{+0.49}_{-0.41}$&$0.96^{+0.04}_{-0.66}$&$0.82^{+1.09}_{-0.66}$&$3.9^{+22.0}_{-3.3}$&8607.37&1.02165\\
C&$1.20^{+2.12}_{-0.50}$&$3.4^{+1.5}_{-0.9}$&$0.35^{+0.65}_{-0.24}$&$0.94^{+0.06}_{-0.62}$&$1.05^{+2.28}_{-0.68}$&$3.3^{+1.7}_{-1.2}$&8604.99&1.02136\\
MC&$1.02^{+0.80}_{-0.49}$&$3.6^{+4.0}_{-1.8}$&$0.44^{+0.56}_{-0.34}$&$0.97^{+0.03}_{-0.67}$&$0.94^{+0.88}_{-0.70}$&$3.5^{+4.1}_{-2.1}$&8610.09&1.02197\\
Sphere&$1.07^{+0.55}_{-0.39}$&$3.6^{+5.4}_{-2.0}$&&&&&8605.82&1.02098\\
$\theta$&$0.96^{+1.26}_{-0.53}$&$2.5^{+10.9}_{-1.8}$&$0.45^{+0.55}_{-0.34}$&$0.75^{+0.25}_{-0.45}$&$0.89^{+1.33}_{-0.74}$&$2.5^{+10.9}_{-2.0}$&8604.31&1.02128\\
X-ray&$1.06^{+1.68}_{-0.43}$&$4.3^{+1.7}_{-2.3}$&$0.41^{+0.58}_{-0.30}$&$0.93^{+0.07}_{-0.61}$&$0.98^{+1.76}_{-0.67}$&$4.2^{+1.8}_{-2.6}$&8605.83&1.02146\\
X-ray $+\theta$&$1.04^{+0.98}_{-0.59}$&$3.9^{+2.0}_{-2.5}$&$0.14^{+0.85}_{-0.03}$&$0.44^{+0.56}_{-0.14}$&$0.73^{+1.29}_{-0.53}$&$3.5^{+2.4}_{-2.4}$&8605.55&1.02143\\
\hline
A2204&&&&\\
Flat&$0.79^{+1.07}_{-0.44}$&$3.6^{+13.9}_{-2.9}$&$0.33^{+0.66}_{-0.24}$&$0.98^{+0.02}_{-0.68}$&$0.69^{+1.18}_{-0.57}$&$3.4^{+14.0}_{-3.0}$&12470.7&1.05594\\
Shaw&$0.72^{+0.54}_{-0.35}$&$4.6^{+11.0}_{-3.1}$&$0.78^{+0.18}_{-0.27}$&$0.87^{+0.13}_{-0.24}$&$0.72^{+0.55}_{-0.37}$&$4.6^{+11.0}_{-3.2}$&12474.3&1.05625\\
Mass&$0.64^{+0.84}_{-0.35}$&$4.2^{+20.0}_{-3.6}$&$0.29^{+0.71}_{-0.19}$&$0.93^{+0.07}_{-0.63}$&$0.54^{+0.94}_{-0.45}$&$4.0^{+20.2}_{-3.6}$&12481.5&1.05686\\
C&$0.79^{+1.07}_{-0.44}$&$3.6^{+2.8}_{-1.2}$&$0.33^{+0.65}_{-0.24}$&$0.98^{+0.02}_{-0.68}$&$0.68^{+1.18}_{-0.51}$&$3.4^{+2.9}_{-1.5}$&12470.7&1.05594\\
MC&$0.64^{+0.76}_{-0.32}$&$4.2^{+5.4}_{-2.5}$&$0.29^{+0.71}_{-0.19}$&$0.93^{+0.07}_{-0.62}$&$0.54^{+0.86}_{-0.40}$&$4.0^{+5.6}_{-2.7}$&12481.5&1.05686\\
Sphere&$0.71^{+0.38}_{-0.26}$&$4.5^{+5.4}_{-2.4}$&&&&&12473.7&1.05584\\
\hline
\end{tabular}
\end{table*} 

\section{Weak Lensing Analysis}
\subsection{Observations}
$R$-band images of Abell 1689, Abell 1835, and Abell 2204, observed on May 29-30 2000 with the MPG/ESO Wide-field Imager are shown in Fig. \ref{fig:plot1}.  Table \ref{tab:table1} lists the positions, redshifts, and background source densities used in the lensing analysis for each of the three clusters.  Details of the observations, shape measurement techniques, and background galaxy selection are given in \cite{clowe01} (herein CS01) for Abell 1689 and in \cite{clowe02} (herein CS02) for Abell 1835 and Abell 2204.  Since those publications, new photometric data were acquired from CFHT observations of Abell 1689 and Abell 1835 (\cite{bardeau05}), supplementing the original $R$-band observations with photometry in $B$- and $I$-bands for A1689 and $V-$ and $I$- bands for A1835. Abell 2204 was already observed in $B$ and $V$ with the WFI in CS02.  This photometric information allows for colour-based cuts to be used in selecting the background galaxy population, in addition to the simple magnitude cuts used for A1689 and A1835 in CS02 and CS01; the colour criteria used for the background galaxies in the field of Abell 1689 are described in \cite{clowe03} (herein C03).  This allows for a much improved selection of background galaxies; most importantly, it allows the removal of faint cluster members, which are indistinguishable from faint background sources in single-band observations.  These cuts were made very conservatively, at a level to exclude all galaxies with $z<0.5$.  For Abell 1689 all galaxies consistent within 2$\sigma$ of being at $z<0.5$ were rejected; for Abell 2204 and Abell 1835 only galaxies within 1$\sigma$ were discarded.  This slightly less stringent standard was required in order to maintain an adequate background source density because the colour fields for those fields were shallower.  The mean lensing redshift of the background galaxy populations is calculated using photo-z information from the COSMOS field (\cite{mobasher07}) by making redshift and magnitude cuts on that catalogue identical to those applied to our cluster lensing catalogues.  It is computed to be $<z_s> = 0.8$ for all three clusters, and this value is used throughout this paper.  Additionally, the A1689 lensing catalogue is updated to include the KSB PSF corrections of \cite{bacon01}, improving the accuracy of the shear measurements.  These corrections were already in place in the CS02 analysis of the other two clusters.

The weak lensing data are taken from an aperture $1.6'<r<18.0'$ in order to exclude the strong lensing regions at the centres of the clusters where $\kappa \sim 1$.  

\begin{figure*}
\centering
\rotatebox{0}{
\resizebox{16cm}{!}
{\includegraphics{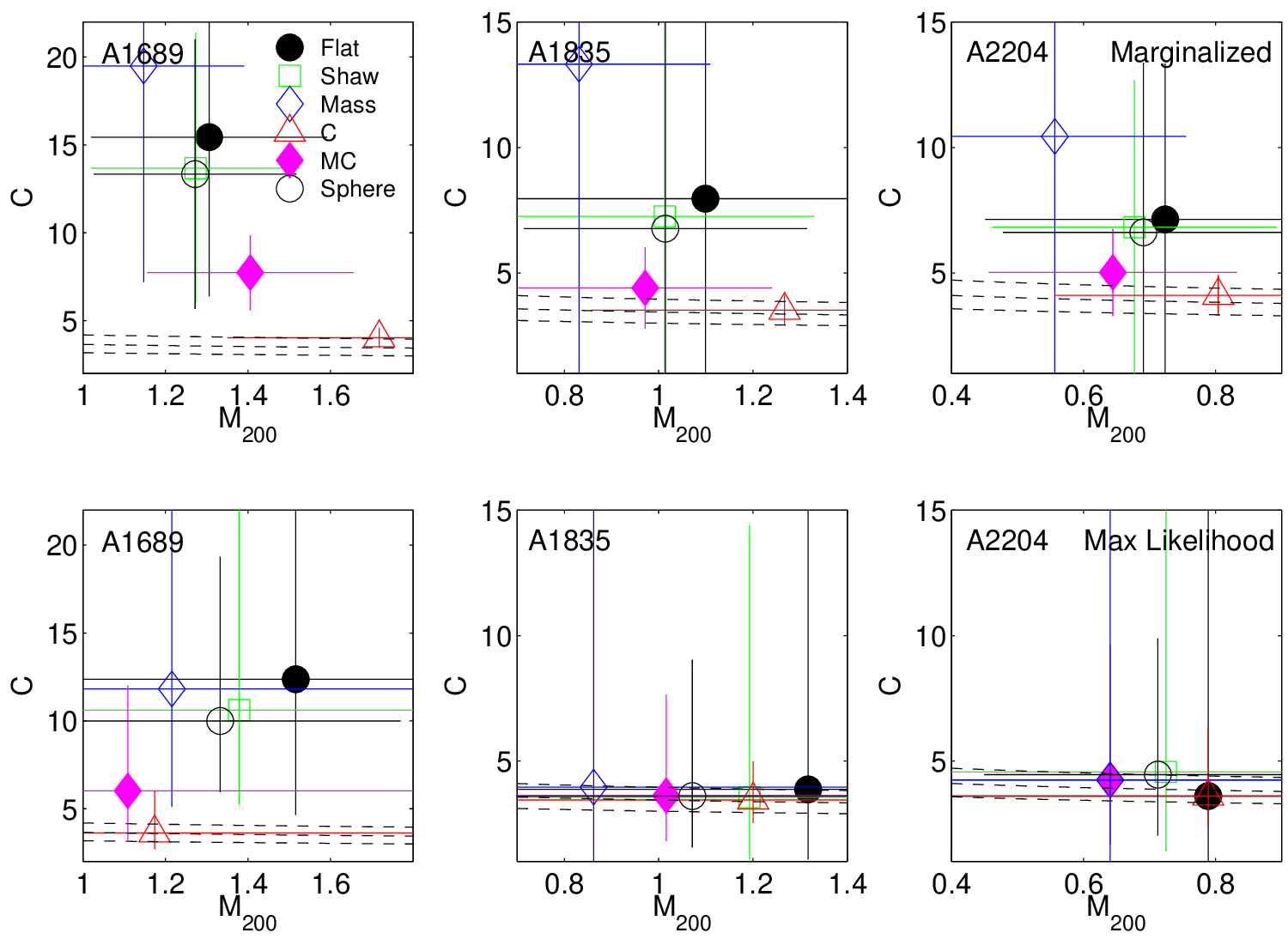}}
}
 \caption{The marginalized mean (top row) and maximum-likelihood (bottom row) $M_{200}$ and $C$ values for A1689, A1835, and A2204 under a flat triaxial prior as well as priors on halo mass, concentration, shape, and orientation derived from $\Lambda$CDM structure formation simulations.  Error bars gives 1$\sigma$ errors on the mean and maximum-likelihood values, and the dotted lines gives the mass-concentration relation predicted in $\Lambda$CDM at $1\sigma$.}
 \label{fig:plot3}
\end{figure*}

\subsection{Weak Lensing by Triaxial Dark Matter Halos}\label{subsec:wl}
Weak lensing distorts the shapes and number densities of background galaxies.  The shape and orientation of a background galaxy can be described by a complex ellipticity $\epsilon^s$, with modulus $|\epsilon^s|=(1-b/a)/(1+b/a)$, where $b/a$ is the minor:major axis ratio, and a phase that is twice the position angle $\phi$, $\epsilon^s=|\epsilon^s|e^{2i\phi}$.  The galaxy's shape is distorted by the weak lensing complex reduced shear, $g=\gamma/(1-\kappa)$, where $\gamma$ is the lensing shear and $\kappa$ the convergence, such that the ellipticity of the lensed galaxy $\epsilon$ becomes
\begin{equation}\epsilon = \frac{\epsilon^s + g}{1 + g^{\ast}\epsilon^s} \approx \epsilon^s + \gamma\label{eq:lens}\end{equation}
in the limit of weak deflections and where $*$ denotes complex conjugation.  Assuming a zero-mean ellipticity for the unlensed population, the expectation values for the lensed ellipticity on a small piece of sky is $<\epsilon> = g \approx \gamma$.  This is the basis for weak lensing analysis in which the shapes of images are measured to estimate the shear profile generated by an astronomical lens; a thorough description of weak lensing is given in \cite{schneider00}.

We calculate the 2D dispersion  in the source ellipticities $\sigma_{\epsilon}=\sqrt{\sigma_{\epsilon 1}^2 + \sigma_{\epsilon 2}^2}$ for each cluster catalogue from the measured 1D dispersion in the radial component of the background galaxy ellipticities measured after all colour and magnitude cuts, to which the gravitational shearing from the cluster contributes negligibly.  It contains both the uncertainties due to the intrinsic noise in galaxy ellipticities and the measurement errors due to sky noise and PSF corrections.  The values for each cluster are give in Table \ref{tab:table1}.

A full parameterisation for a triaxial NFW halo is given by \cite{jing02} (herein JS02).  They generalise the spherical NFW profile to obtain a 
density profile
\begin{equation}
\rho(R) = \frac{\delta_c \rho_c(z)}{R/R_s(1 + R/R_s)^2}
\label{eq:3axrho}
\end{equation}
where $\delta_c$ is the characteristic overdensity of the halo, $\rho_c$ the critical density of the Universe at the redshift $z$ of the cluster, $R_s$ a scale radius, $R$ a triaxial radius 
\begin{equation}
R^2 = \frac{X^2}{a^2} + \frac{Y^2}{b^2} + \frac{Z^2}{c^2},\textrm{         }(a\leq  b \leq  c = 1),\label{eq:3axR}\end{equation}
and $a/c$ and $b/c$ the minor:major and intermediate:major axis ratios, respectively.  In a  different choice from JS02 we define a triaxial virial radius $R_{200}$ such that the mean density contained within an ellipsoid of semi-major axis $R_{200}$ is $200\rho_c$ such that the concentration is
\begin{equation}C = \frac{R_{200}}{R_s},\label{eq:3axC}\end{equation}
the characteristic overdensity is
\begin{equation}\delta_c = \frac{200}{3} \frac{C^3}{\log (1+C) - \frac{C}{1 + C}},\label{eq:3axdelta}\end{equation}
the same as for a spherical NFW profile, and the virial mass is
\begin{equation}M_{200} = \frac{800\pi}{3}abR_{200}^3\rho_c.\label{eq:3axM200}\end{equation}
A more detailed description of this triaxial parameterisation and its benefits is given in \cite{corless07}.

To allow comparison to some past lensing and numerical studies employing spherical models, we define an effective spherical virial mass and concentration which we give in addition to the fully triaxial quantities when quoting marginalized and maximum-likelihood parameters.  The effective spherical virial radius is defined as the radius $r_{200}$ at which the mean density within a sphere of that radius is 200 times the critical density, and the effective spherical virial mass the mass within that sphere: 
\begin{equation} m_{200} = (800\pi /3) r_{200}^3\rho_c.\end{equation}
We further define the effective spherical concentration $C_{sph}$ of a triaxial halo as the ratio of the effective spherical virial radius to the geometric mean of the triaxial scale radii $r_s = R_s(abc)^{1/3}$:
\begin{equation}C_{sph} = r_{200}/r_s.\end{equation}
Throughout this work we always directly fit the triaxial mass and concentration, then calculate the effective spherical values from those fitted triaxial models.  For a spherical model, the triaxial and effective spherical parameterizations are identical: $m_{200} = M_{200}$ and $C_{sph} = C$.  We consistently quote masses and concentrations derived using spherical models in our own work and that of other authors as the more general $M_{200}$ and $C$.

The triaxial halo is oriented with respect to the observer by angles $\theta$, the orientation angle between the major axis of the halo and the observer's line-of-sight, and $\phi$; randomly oriented halos are distributed uniformly in $\phi$ and sin$\theta$.  

The lensing properties of a triaxial halo, expressed as $\gamma$ and $\kappa$, are determined by performing integrals over the spherical NFW convergence and its derivatives. They are derived in \cite{oguri03} (herein OLS), and we summarised and extended some of that work in \cite{corless07}; it is included here in Appendix \ref{sec:appa}.

\subsection{MCMC Fitting Method}
A more detailed description of the MCMC method for fitting triaxial NFW halos is given in CK08.  MCMC methods employ a ``guided'' random walk that returns a sample of points representative of the posterior probability distribution; the probability of a certain region of parameter space containing the true model is directly proportional to the density of points sampled in that region.  From the distribution of sample points the full posterior probability distribution is obtained, which is easily and directly marginalized over to obtain fully marginalized mean most-probable parameter estimates for all parameters. Such methods are particularly valuable in under-constrained or highly-degenerate fitting problems such as this one of fitting 3D triaxial models to 2D lensing data.  By contrast, with such weak constraints a maximum likelihood approach is both impractical and of very limited scientific value; the posterior distribution will be very flat with significant degeneracies giving poorly constrained maximum likelihood values.  By exploring the full posterior probability distribution, this MCMC method allows the derivation of parameter estimates (and their accompanying errors) that account for the true uncertainties when fitting parametric models to lensing data.

To implement the method, we must define the posterior probability function, defined in Bayesian statistics as
\begin{equation}p(\pi|\theta) = \frac{p(\theta|\pi)p(\pi)}{p(\theta)}\end{equation}
where $p(\theta|\pi)$ is the likelihood $\mathcal{L}$ of the data given the model parameters (the standard likelihood), $p(\pi)$ is the prior probability distribution for the model parameters (e.g. a distribution of axis ratios drawn from simulations), and $p(\theta)$ is a normalising factor called the {\it evidence}, of great value in comparing models of different classes and parameter types, but computationally expensive to calculate and unnecessary for the accurate exploration of the posterior distribution.  We define the log-likelihood function in the standard manner for weak lensing following \cite{schneider00} and \cite{king01}
\begin{equation}\ell_{\gamma} = -\ln \mathcal{L} = -\sum_{i=1}^{n_{\gamma}}\ln  p_{\epsilon}(\epsilon_i|g(\vec{ \mathcal{\theta}_i}; \Pi)).\label{eq:like}\end{equation}
where the reduced shear $g$ is calculated using the triaxial convergence and shear of Equations \ref{eq:gammakappa}, $n_{\gamma}$ is the number of background galaxies used in the analysis, and $\Pi$ is a six-element vector of the parameters defining the model: triaxial virial mass $M_{200}$, concentration $C$, minor axis ratio $a$, intermediate axis ratio $b$, and two orientation angles $\theta$ and $\phi$.   We note that while the angles are defined over a range $0<\theta <\pi$ and $0< \phi < 2\pi$, because of the elliptical nature of the projected density contours, they give rise to unique lensing profiles only over the range of $0<\theta <\pi/2$ and $0< \phi < \pi$. The prior probability distribution, already mentioned in the introduction, will be discussed in detail in the next section.

In our MCMC sampler we use the covariance matrix of an early run to sample in an optimised basis aligned with the degeneracies of the posterior.  We tune the step sizes of the sampler to achieve an average acceptance rate of 1/3 in each basis direction, run three independent MCMC chains, started at randomly chosen positions in parameter space, for each lens, and sample the distribution space until the standard var(chain mean)/mean(chain var) indicator is less than 0.15, strongly indicating chain convergence (\cite{lewis02}).  We utilise the GetDist package from the standard CosmoMC (\citealt{lewis02}) distribution to calculate convergence statistics, confidence contours, and marginalized and maximum-likelihood parameter estimates.  We employ 60 bins in the Gaussian smoothing of the contours; choosing such a large number leads to noisier contours, with more small islands of confidence and irregularly shaped confidence regions, but is the level at which the size of the contours is no longer sensitive to the bin number.  If a lower number is employed, the contours look much smoother but grow significantly larger leading to overestimates of the confidence regions.

Throughout the paper we quote both marginalized and maximum-likelihood parameter values.  It is important to note that the maximum-likelihood values returned by MCMC methods such as this do not correspond to the true likelihood maximum -- no minimization to find the exact peak of the likelihood distribution is attempted because in our six-parameter, degenerate triaxial model such peaks are poorly constrained.  Instead, the quoted maximum likelihood values are the parameter values at the point with maximum likelihood in the MCMC sample.  These will correspond to values close to those at the peak of the likelihood distribution, and are included to demonstrate the offset that may occur between the maximum likelihood values and the marginalized mean parameter values.  This offset occurs because there are significant asymmetric degeneracies between the triaxial parameters and the errors on those parameters are highly non-Gaussian.  The two types of ``best'' parameter values reflect slightly different information about the lensing cluster: the maximum-likelihood value gives the one point at which the very best fit is attained, while the marginalized values give the region of parameter space where most probability resides.  The marginalized statistics are more robust than the maximum-likelihood values, especially for an underconstrained problem such as this in which the degeneracies are significant and the errors large; this is reflected in the much larger error bars on the maximum-likelihood values.  We will therefore treat the marginalized parameter values as the ``best'' parameter estimates for our clusters, but quote maximum-likelihood values as well to provide a better sense of the one best model and to allow for more direct comparisons with previous studies which typically quote maximum-likelihood or minimum $\chi$-squared parameter estimates and errors.  The errors quoted for the marginalized mean parameter estimates are the 1$\sigma$ standard deviation; those for the likelihood estimates are the one-sided $68\%$ 1$\sigma$ limits.

\subsection{Prior Probability Functions}\label{sec:priors}
As noted earlier, because lensing is a fundamentally 2D phenomenon, models of 3D structures fit using lensing data alone -- whether strong, weak, or flexion -- are fundamentally underconstrained.  It may therefore be of interest to further constrain the triaxial models with prior probability functions derived from theoretical predictions or complementary observations.  Several of these functions were discussed in CK08; some of those are included here, and others in this paper are improvements or additions to the set presented in that work.  Importantly, CK08 showed that the triaxial fitting method applied to a population of triaxial halos using a relatively accurate prior on the distribution of triaxial axis ratios returns statistically accurate and robust estimates of parameter values across that population.  However, given our limited knowledge of the universe (thus our desire to model dark matter halos in the first place) the choice of prior is not universal or obvious, and the choice must be carefully made with respect to the particular problem at hand.  For example, when testing the mass-concentration relation that dark matter halos are predicted to exhibit in CDM (\cite{navarro97}; \cite{neto07}), adopting a prior based on that relation would be inappropriate.  However, in another problem, when modelling a large population of dark matter halos to calculate a mass function, a loose mass-concentration prior may be an appropriate way to incorporate existing knowledge of the cluster and group population into the model.  Similarly, a strict prior on triaxial axis ratios may be well-justified when applied across a large population, but may be a poor choice when looking at a single strong-lensing cluster likely to be an outlier in the triaxial distribution.

The prior probability functions examined in this paper are described below, and plotted in Figure \ref{fig:plot2}.  All priors but the most general are fundamentally grounded in CDM, whether directly from CDM structure formation simulations or indirectly from work employing NFW profiles derived from those structure formation simulations.
\begin{itemize}
\item {\it Flat}: The Flat prior is a very general prior on the shape and orientation of triaxial halos.  It is defined (slightly more broadly than in CK08) to have probability $p=1$ when
\begin{eqnarray}
0.3\leq &b&\leq 1.0\textrm{   and}\nonumber\\
0.3\leq &a/b&\leq 1.0,\end{eqnarray}
and $p=0$ elsewhere.  It excludes models with extremely small axis ratios because they are not expected in nature, and their inclusion slows convergence of the MCMC chains significantly. It also includes a prior on the orientation angles of the triaxial halo corresponding to random orientation with respect to the line-of-sight:
\begin{eqnarray}
p(\theta) &=& 0.5\sin\theta;\nonumber\\
p(\phi) &=& \frac{1}{2\pi}.\end{eqnarray}
This prior is the most basic employed in this paper, and it is applied in every case -- all other priors are implemented on top of it.  
\item {\it Shaw}: The Shaw prior is a CDM prior on the triaxial axis ratios, employing the axis ratio distributions found in the CDM structure formation simulations of \cite{shaw06}.  These distributions are fit using polynomials; these are plotted in the first panel of Figure \ref{fig:plot2} and given explicitly in Appendix \ref{sec:appb}.
\item {\it Mass}: The Mass prior is a CDM prior on halo mass, taken from the mass function fit to the $\Lambda$CDM structure formation simulations of \cite{evrard02}.  The differential number density of dark matter halos as a function of mass and redshift is given by 
\begin{equation}n(M,z) = \frac{A\bar{\rho}_m(z)}{M}\alpha_{eff}(M)\exp\left[-|\ln\sigma^{-1}(M)+B|^{\epsilon}\right]\end{equation}
where $\alpha_{eff}$ is the effective logarithmic slope, $\sigma^2(M)$ is the variance of the density field smoothed on scales enclosing mass $M$ at the mean density $\bar{\rho}_m(z)$, and $A$, $B$, and $\epsilon$ are free parameters fitted for a $\Lambda$CDM universe in Evrard et al.  This prior favours lower mass halos, as seen in the second panel of Figure \ref{fig:plot2}.
\item {\it Concentration}: The Concentration (C) prior is a CDM prior relating halo concentration to mass, taken from the mass-concentration relation derived by \cite{neto07} from the simulated groups and clusters in the Millennium simulation:  
\begin{equation}C = \frac{5.26}{1+z}\left(\frac{M_{200}}{10^{14} h^{-1} M_{\odot}}\right)^{-0.1},\label{eq:M-C}\end{equation}
with a log-normal scatter
\begin{equation}p(\log C) = \frac{1}{\sigma \sqrt{2\pi}}\exp\left[-\frac{1}{2}\left(\frac{\log C - <\log C>}{\sigma}\right)^2\right]\label{eq:cdisp}\end{equation}
where $<\log C>$ is calculated via Eq. \ref{eq:M-C} for a given $M_{200}$ and the dispersion $\sigma$ is taken to be 0.09 for masses less than $10^{15}$ M$_{\odot}$ and 0.06 for masses greater than that threshold, in rough agreement with the results of \cite{neto07}. Examples of this prior at different values of $M_{200}$ are plotted in the third panel of Fig. \ref{fig:plot2}.
\item {\it Mass + Concentration}: The Mass-Concentration (MC) prior combines the predictions of CDM for mass and concentration.  It employs the Evrard mass function of the Mass prior, and a looser form of the mass-concentration relation of the C prior, plotted as the ``Wide'' prior in the third panel of Fig. \ref{fig:plot2}.  The dispersion of the log-normal probability distribution of concentrations at a given mass $\sigma$, as defined for the $C$ prior in Eq. \ref{eq:cdisp}, is taken to be 0.18.  That value, compared to those derived from simulations and used in the $C$ prior above, is twice that for lower mass halos and three times that of higher mass halos.  This increased flexibility makes this a more general (though still CDM-based) physical prior, that favours generally CDM-like halos but does not overpower the preferences of observed lensing data.
\item {\it Sphere}: The Spherical prior sets both axis ratios to unity, $a=b=1$.  It is included for comparison to the spherical fits typical of most lensing analyses.
\end{itemize}

\begin{figure*}
\centering
\rotatebox{0}{
\resizebox{16cm}{!}
{\includegraphics{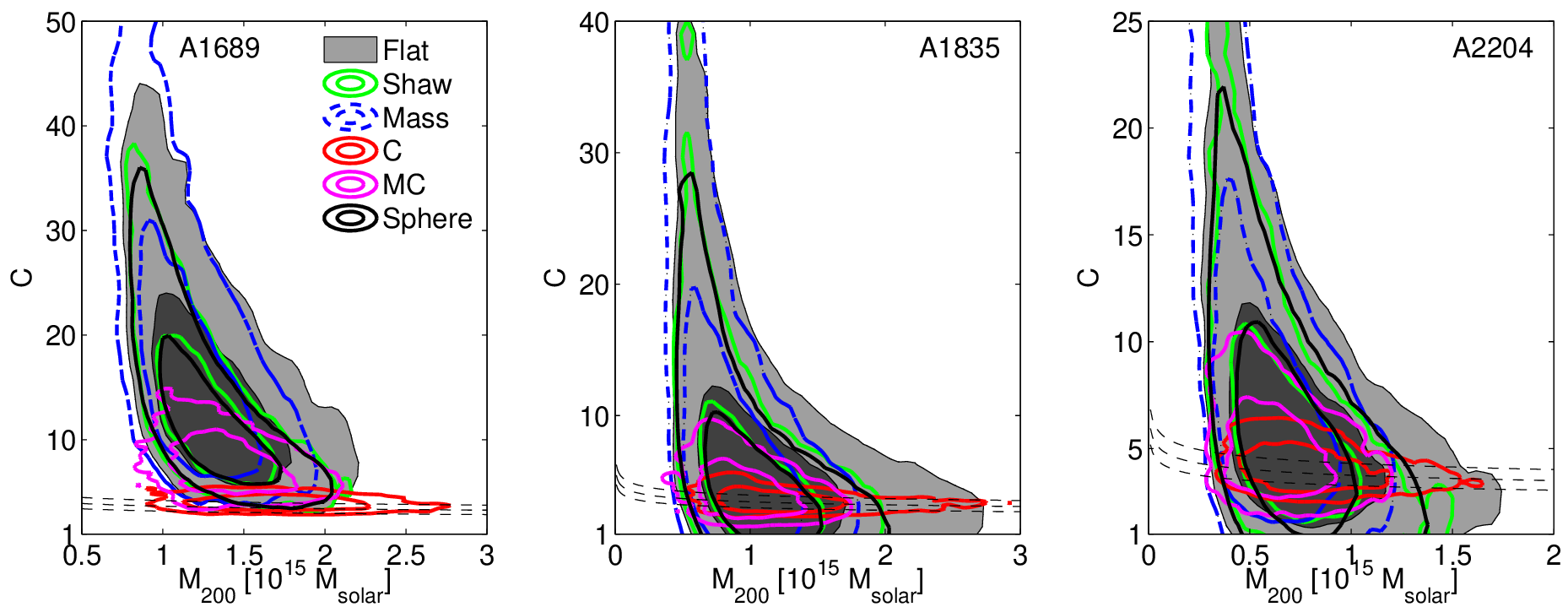}}
}
 \caption{The 2D marginalized $C-M_{200}$ 68 and 95 \% confidence contours for the triaxial NFW fit to A1689, A1835, and A2204 under a Flat triaxial prior as well as priors on halo mass, concentration, shape, and orientation derived from $\Lambda$CDM structure formation simulations.  All priors are described in Sec. \ref{sec:priors}.}
 \label{fig:plot4}
\end{figure*}

\begin{figure*}
\centering
\rotatebox{0}{
\resizebox{16cm}{!}
{\includegraphics{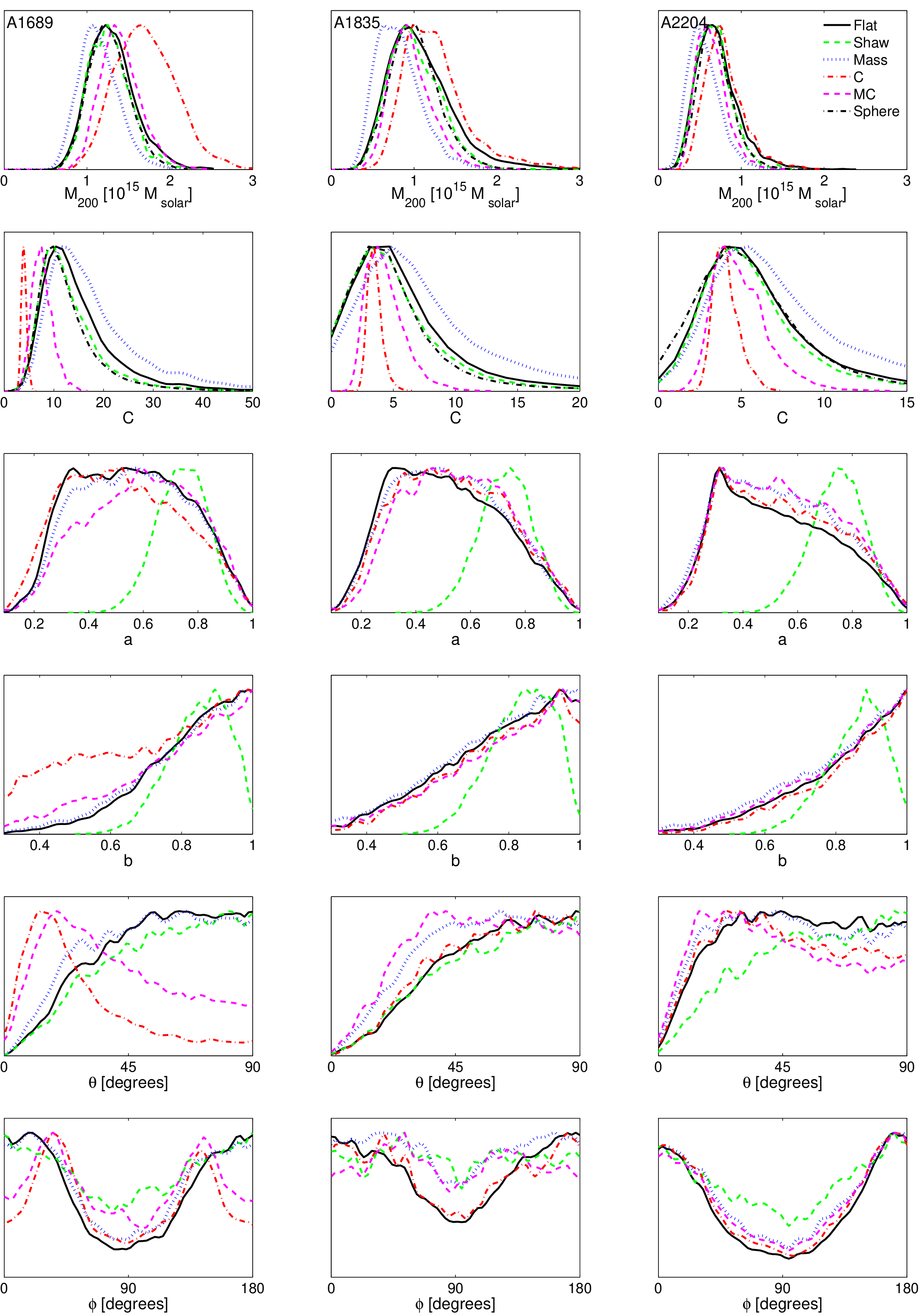}}
}
 \caption{The 1D marginalized probability distributions for all six parameters of the triaxial NFW fit to A1689, A1835, and A2204 under a Flat triaxial prior as well as priors on halo mass, concentration, shape, and orientation derived from $\Lambda$CDM structure formation simulations.  All priors are described in Sec. \ref{sec:priors}.}
 \label{fig:plot5}
\end{figure*}

\section{Results}
Under each of these priors we fit the lensing signal of each of the three clusters with a single triaxial NFW halo using the MCMC method, taking lensing data from an aperture $1.6'<r<18.0'$.  We treat all background galaxies as though they are located on a sheet at redshift $<z_s>=0.8$; the low redshifts of the lensing clusters justify this simplifying assumption, as only for higher redshift lenses that are in the heart of the redshift distribution is the distribution of source redshifts important \citep{seitz97}.

Tables \ref{tab:table2} and \ref{tab:table3} give the mean marginalized and maximum-likelihood parameter values with 1$\sigma$ errors for all three clusters under each of the priors listed in Section \ref{sec:priors}.  The corresponding effective spherical parameters are also given, as well as the $\chi^2$ and reduced $\chi^2$ of the maximum-likelihood triaxial fit to the lensing data.  Figure \ref{fig:plot3} plots the mean marginalized and maximum-likelihood $M_{200}$ and $C$ values, and the dotted lines show the predicted CDM mass-concentration relation with $1\sigma$ errors.  Figure \ref{fig:plot4} gives the $1$- and $2\sigma$ (68 and 95 percent) $M_{200}-C$ confidence contours for each cluster and prior, again with the CDM mass-concentration relation overplotted in dashed lines.  Figure \ref{fig:plot5} gives the marginalized posterior probability distributions of all six triaxial parameters for each cluster and prior.

Priors are clearly very important, as the parameter mean values, contours, and 1D probability distributions shift significantly depending on the prior applied.  The impact of each prior differs between the three clusters, which is excellent evidence that the priors interact with the lensing data rather than overpower it. Examining the marginalized and maximum-likelihood parameter estimates, the 1D marginalized distributions, and the $M_{200}-C$ contours in combination gives insight into the way the various priors function.  
\begin{itemize}
\item {\it Shaw}: In every case the Shaw prior significantly changes the posterior probability distributions for the axis ratios, as expected since in its absence the constraints on the axis ratios are very weak, as the projected ellipticity and orientation angle constrain two axis ratios and two 3D orientation angles.  However, it is important to note the case of Abell 2204, where even with only a Flat prior the lensing data proves very capable of constraining the minor axis ratio (this signifies a large ellipticity on the sky), and the Shaw prior almost completely overrides the data when it is applied.  Thus, while the Shaw prior was shown in CK08 to be quite good for determining the {\it statistics} of a large population of dark matter halos, it may suppress real information in the lensing data when interpreting individual clusters.  The large shift in the posterior probability distribution of the axis ratios forced by the Shaw prior leads to changes in the preferred orientation angles -- see for example that in Abell 2204 line-of-sight orientation angles close to $\theta=0$ are no longer favoured, and that the 1D distributions for $\phi$ flatten considerably for all three clusters.  The impacts on the mass and concentration estimates are relatively small, and generally toward lower values.
\item {\it Mass}: The Mass prior prefers lower masses; in every case this leads to a movement of the $M_{200}-C$ contours to lower masses and higher concentrations, where the higher concentrations serve to boost the lensing signal to counter the reduction in mass.  In all cases the $\theta$ distributions move towards more lensing efficient line-of-sight orientations at $\theta=0$, serving also to counter the decreases in mass induced by the prior.
\item {\it C}: The concentration prior functions very differently for the different clusters.  In Abell 1689, where the contours under the Flat prior strongly favour concentrations higher than expected in CDM, and contain the CDM mass-concentration contours only barely at $2\sigma$, the imposition of the C-prior completely dominates the lensing signal and shifts the contours dramatically to lower concentration values.  By contrast, in the case of A2204 the $C$ prior leads to only a small decrease in the marginalized concentration and no change at all in the maximum likelihood parameters, as the contours under the Flat prior are already consistent with the mass-concentration relation.  Abell 1835 falls between the other two; the Flat prior favours somewhat (theoretically) high values of $C$, and the imposition of the $C$ prior leads to a small shift of the $M_{200}-C$ contours to slightly lower concentration values.  In every case, the marginalized mass moves to counter the prior-induced shift in $C$, moving to higher (lower) masses as $C$ decreases (increases).
\item {\it MC}: As expected, this combination of the mass and (looser) concentration priors combines the effects of its components: in every case the marginalized mean parameter values lie in-between those of the individual Mass and $C$ priors, as seen in Fig. \ref{fig:plot3}.  The $MC$ prior generally moves the parameter estimates of the Flat prior towards the CDM mass-concentration relation, but weakly: the $M_{200}-C$ confidence contours under the MC prior are almost entirely contained within those of the Flat prior in all cases.  Still, however, the shift in mass and concentration parameters is quite dramatic for the outlier Abell 1689.  Also note that the 1D distributions for the axis ratios $a$ and $b$ and orientation angles $\theta$ and $\phi$ typically fall between those of the two components, but usually follow those of the Mass prior more closely.
\item {\it Sphere}: The spherical prior generally prefers lower masses and concentrations to the Flat triaxial prior; this is clearly seen in the $M_{200}-C$ contours, as the Flat contours are enlarged from their spherical counterparts in all directions but preferentially towards higher masses and concentration.  This is because there are more inefficient lensing orientations for a triaxial halo than there are efficient ones. Thus, as argued already in CK08, halo triaxiality cannot explain a whole population of apparently over-concentrated halos, though it can quite easily account for some individual cases.  Under this prior the estimated error contours on the parameters are much smaller than those under the Flat or Mass priors, but not as constraining as the concentration priors $C$ and $MC$.
\end{itemize}

The varied results for each of the clusters under the different priors indicate that the priors are not generally dominating the lensing signal.  The stringent Shaw and $C$ priors are seen to sometimes work strongly against the data, but this is less of a concern because the impacts are so obvious -- for example, the Shaw prior may be valid for Abell 1689, for example, where the marginalized 1D distributions for $a$ and $b$ are entirely enclosed by their counterparts under the Flat prior, but it is clearly not for Abell 1835 and especially Abell 2204 where they sit mostly outside the more general distributions.  Thus the Shaw prior is often not a good choice for studies of individual clusters; however, it may still be the best choice for statistical studies, as it was shown to perform very well across populations in CK08.  Similarly, the $C$ prior applied to Abell 1689 shifts the contours dramatically outside of those of the Flat prior, indicating it is a bad choice in this problem.  Though the Mass prior does not behave so badly, it is an unbalanced expression of the CDM model, favouring lower masses while pushing up concentration.  Generally, the $MC$ prior is the best behaved, weakly reflecting true prior expectations for any galaxy cluster, exerting a balanced influence on both mass and concentration while never entirely over-riding the data.

In the case of Abell 1689 we find evidence supporting a concentration significantly higher than that predicted by CDM simulations.  Every prior but the $C$ and $MC$ priors, which explicitly forces the solution towards the predicted mass-concentration relation, gives marginalized and maximum-likelihood values of concentration well above the $1\sigma$ CDM mass-concentration corridor.  In the $M_{200}-C$ contours the weak lensing data of A1689 are seen to be at best marginally compatible with the CDM mass concentration relation at $2\sigma$ for all but the $C$ and $MC$ priors.  

The reduced $\chi^2$ values are slightly greater than unity for all three clusters under all priors, indicating the NFW density profile is a good fit to the lensing data; the values are expected to be slightly larger than unity, as they are, because our models do not fit any substructure.

\begin{figure*}
\centering
\rotatebox{0}{
\resizebox{16cm}{!}
{\includegraphics{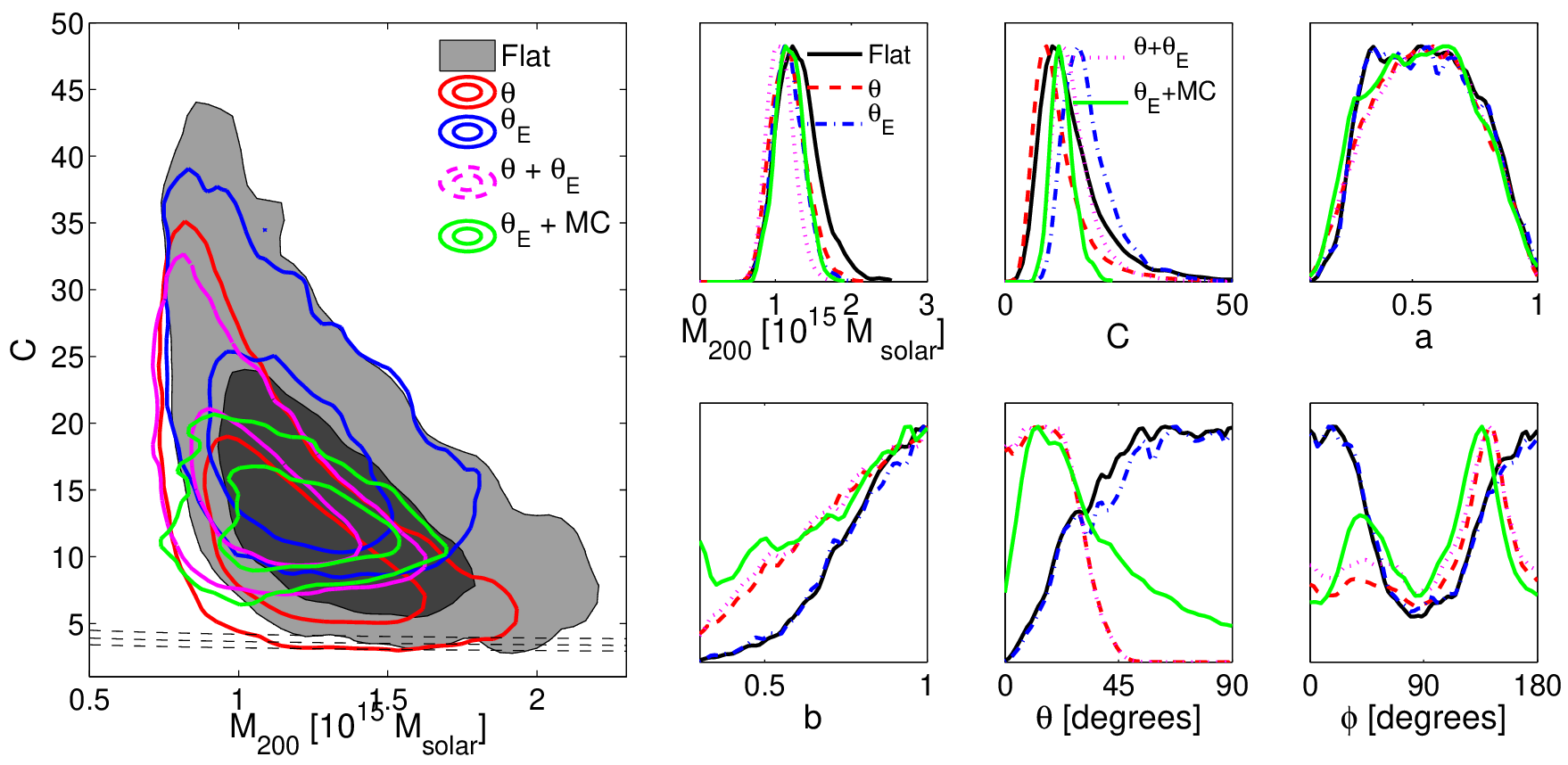}}
}
 \caption{The left-hand panel plots the 2D marginalized 68 and 95 \% confidence contours in the $C-M_{200}$ plane for A1689 under the Flat, $\theta$, $\theta_E$, and $MC$ priors, described in the text.  The dotted lines show the region of parameter space consistent at $1\sigma$ with the mass-concentration relation predicted in $\Lambda$CDM.  The right-hand panels give the corresponding 1D marginalized probability distributions for all six triaxial parameter under each prior.}
 \label{fig:plot6}
\end{figure*}

\subsection{Cluster-Specific Priors}\label{sec:extpriors}
In addition to these general priors, if independent constraints on mass and concentration  exist for a cluster from X-ray, strong lensing, or dynamical data, it may be desirable to combine them with the weak lensing analysis to derive joint constraints.  In this section we examine what external data is available for each of these clusters and how it may be combined to tighten constraints on the triaxial NFW.  Care is necessary in doing so, for in some cases the very assumptions of sphericity we are trying to avoid may be implicit in the parameter values derived from other methods, and so combination with these results may be detrimental to the accuracy of our final mass and concentration constraints.  
\begin{figure}
\centering
\rotatebox{0}{
\resizebox{6cm}{!}
{\includegraphics{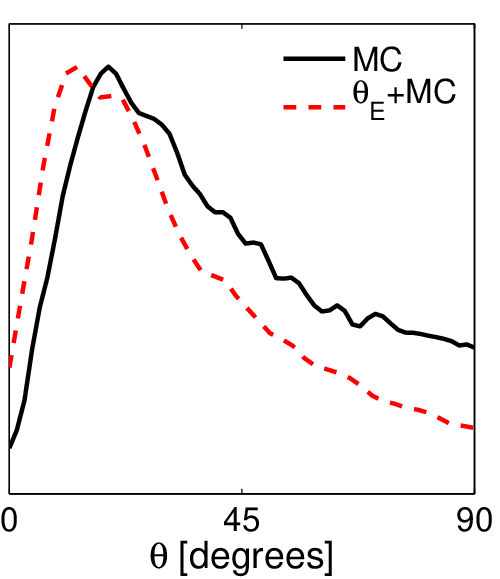}}
}
 \caption{The 1D marginalized distribution for the triaxial line-of-sight orientation angle $\theta$ For Abell 1689 under the general CDM $MC$ prior alone (plotted in solid black) compared with that under the $MC$ prior in combination with the observational Einstein radius $\theta_E$ prior (plotted in dashed red).  The addition of the Einstein radius constraint to the general CDM prior favors more line-of-sight oriented models, supporting the predictions of \citet{oguri08}.}
 \label{fig:plot7}
\end{figure}

\subsubsection{Abell 1689}
Abell 1689 presents a wealth of strong lensing features, and these have been modelled extensively (e.g. \cite{miralda95}, \cite{broadhurst05a},\cite{limousin07}).  Though many studies publish NFW mass and concentration parameters, spherical symmetry is always assumed in their derivation, and so direct application of these constraints as a prior on our triaxial model is not sensible.  However, a model independent quantity that is very reliably constrained by the strong lensing work is the Einstein radius $\theta_E$ (the location of the lensing critical curves where giant arcs appear); for Abell 1689 $\theta_E$ is consistently measured to be $45''$ for a spectroscopically constrained mean source redshift $z_s =1$.  There are two possible ways to apply this constraint:
\begin{itemize}
\item \cite{oguri08} demonstrate that galaxy clusters with relatively large Einstein radii such as Abell 1689 are strongly biased towards line-of-sight orientations where $\theta$ is close to zero.  They express this bias as the fraction of halos with Einstein radii above a cutoff value for which $|\cos\theta| > 0.9$: for the Einstein radius of Abell 1689 this fraction is approximately $80\%$.  We give this probability distribution  analytic form with a Gaussian peaked at $\theta=0$ with a dispersion that gives the correct fraction of systems with $|\cos\theta| > 0.9$:
\begin{equation}p(\theta) \propto \frac{1}{\sigma_\theta \sqrt{2\pi}}\exp\left[-\frac{\left(|\cos\theta| - 1\right)^2}{2\sigma_\theta^2}\right]\label{eq:tprior}\end{equation}
where $\sigma_\theta = 0.115$.  This prior, effectively a CDM prior on the orientation of strong lensing halos, which we will call the $\theta$ prior, is plotted in the fourth panel of Figure \ref{fig:plot2}, compared with the distribution for unbiased random halo orientations.
\item {\it Einstein Radius}: Because the mean source redshift is well-determined for the Einstein radius of Abell 1689, it is also possible to more directly constrain the weak lensing models, demanding the models give an Einstein radius for $z_s=1$ of $45''$, allowing for a $10\%$ error in the measurement of $\theta_E$ due to uncertainties in the redshift of the lensed source or the impact of baryons at the very centre of the cluster ($\sigma_{\theta_E} = 0.1\theta_E$):
\begin{equation}p(\Pi) \propto \frac{1}{\sigma_{\theta_E} \sqrt{2\pi}}\exp\left[-\frac{\left(\theta_E - \theta_{E_{true}}\right)^2}{2\sigma_{\theta_E}^2}\right].\end{equation}
\end{itemize}

The 2D $M_{200}-C$ confidence contours and 1D marginalized parameter distributions under both of these priors are shown in Figure \ref{fig:plot6}, as well as under both priors in combination; the resulting marginalized and maximum-likelihood parameter estimates are in Tables \ref{tab:table2} and \ref{tab:table3}.

The application of the $\theta$ prior significantly shifts the contours to lower values of mass and concentration; this is expected because it explicitly favors more efficient lensing orientations.  It also noticeably increases the probability of more extreme values of the intermediate axis ratio $b$.  In the line-of-sight orientations favored by the $\theta$ prior, circular symmetry in projection is achieved as $a$ and $b$ become close to equal in value, so for the relatively circular A1689 $b$ will follow $a$ more closely under this prior.  Lower values of the axis ratios are favored because, as shown in \citet{corless07} and \citet{oguri08}, small axis ratios give the most efficient lensing for halos in line-of-sight orientations.  As expected, the distribution of $\theta$ values shifts towards zero, and this tighter constraint on one orientation leads to a corresponding tightening of the distribution of its counterpart $\phi$.

The Einstein radius $\theta_E$ prior shifts the $M_{200}-C$ contours differently, towards slightly higher concentrations and lower masses.  The move away from higher masses suggests that increasing concentration is a more efficient way of boosting the Einstein radius than increasing the mass.  This preference for higher concentrations as opposed to higher masses also is consistent with the fact that it is the center of the cluster that determines the strong lensing behavior as opposed to the overall profile.  This is important because it is at the cluster center that the impact of baryons, whether in the central cluster galaxies or the intracluster medium, are expected to be most important.  Thus, this more stringent Einstein radius prior may not be well-suited for application to the fitting of a dark-matter only profile, as it may enhance the impact of baryons at the dense cluster center.  However, \citet{meneghetti03b} and \citet{wambsganss08} have both shown the impact of the central cluster galaxy on the strong lensing cross section to be significantly scale dependent: it may increase the cross section by up to $100\%$ for small separation systems, but has a negligible impact for large separation systems.  Thus, for a system such as A1689 with a very large Einstein radius, the contribution of baryons to the size of the Einstein radius is predicted to be small.

We also combine the two strong-lensing priors, applying the orientation bias $\theta$ and the Einstein radius $\theta_E$ prior simultaneously.  As might be expected, the resulting $M_{200}-C$ contours fall between those of the two component priors, picking out the lowest mass and concentration halos that generate an Einstein radius as large as that observed for the cluster.  This is the most physically representative of the strong lensing priors: it combines both our expectations for the orientation of the triaxial halo given its role as one of the strongest lensing clusters in the universe with the demand that the fitted models adequately reproduce the observed strong lensing signal.  Under it, A1689 is not consistent with the CDM mass-concentration at over 2$\sigma$. However, as with the $\theta_E$ prior applied alone, the potential impact of baryons at the cluster center is not accounted for. 

As a further test, we combine the observational Einstein radius prior $\theta_E$ with the general CDM $MC$ prior.  The results complete a consistent picture: adding an additional CDM constraint on the mass and concentration to the observational $\theta_E$ prior shifts the distribution of the orientation angles to strongly favor orientations near the line-of-sight ($\theta$ small), thus recovering the alternate CDM prediction, as implemented in the $\theta$ prior, of a strong orientation bias toward low $\theta$ values.  The observational Einstein radius prior increases the preference for low $\theta$ values compared to that under the $MC$ prior alone, as shown in Figure \ref{fig:plot7}; this provides observational support for the link between large Einstein radii and line-of-sight triaxial orientations in a CDM universe predicted by \cite{oguri08}.

Under the $\theta$ prior alone, the 2$\sigma$ $M_{200}-C$ confidence contours enclose the predictions of CDM for the mass-concentration relation.  With the addition of the $\theta_E$ prior, however, the 2$\sigma$ $M_{200}-C$ confidence contours pass far above the predicted CDM mass-concentration corridor; the combined $\theta_E +MC$ prior also excludes the predicted mass-concentration relation.  Note that under the $\theta$ prior both the triaxial mass and concentration estimates, and under the combined prior the triaxial mass only, are lower than their counterparts under the Spherical prior; this indicates that use of the spherical model may indeed lead to false conclusions regarding the place of A1689 within the CDM paradigm.  Due to remaining uncertainty regarding the potential importance of baryons at the cluster center, especially if A1689 has undergone a recent merger as the dynamical and X-ray data suggest it may have, we hesitate to quote results employing the Einstein radius prior as the definitive triaxial model for the cluster.  We therefore take as the best current constraint on the triaxial NFW model of A1689 the values derived under the less-restrictive strong lensing orientation $\theta$ prior alone.  These are $M_{200} = (1.18 \pm 0.23)\times 10^{15}$ M$_{\odot}$ and $C=12.2\pm 6.7$ ($M_{ES} = (1.16^{+0.25}_{-0.30})\times 10^{15}$ M$_{\odot}$ and $C_{sph}=12.1\pm6.8$), which are marginally consistent at $2\sigma$ with the prediction of CDM.  However, the size and shape of the $M_{200}-C$ confidence contours that stretch towards very high values of concentration while only grazing the predicted mass-concentration corridor, combined with the complete exclusion of that corridor when the triaxial model is fit under the combined $\theta + \theta_E$ and $\theta_E + MC$ priors, indicates that triaxiality alone does not fully explain A1689's unusual characteristics: comparison of it with the other two clusters in this study shows that A1689 is still unusual for the high masses and concentrations its lensing data favor.

Abell 1689 has also been observed in the X-ray using both Chandra and XMM-Newton (\cite{xue02}; \cite{andersson04}; \cite{lemze08}).  However, X-ray mass models require the assumption of both spherical symmetry and hydrostatic equilibrium.  Generally, because the gravitational potential traced out by the X-ray gas is always rounder than the underlying density distribution, X-ray models assuming spherical symmetry are less likely to be strongly affected by triaxiality and elongation of the cluster along the line-of-sight (see \cite{kassiola93}; \cite{russell08}).  However, the strong orientation bias of \cite{oguri08} indicates Abell 1689 is likely to be very elongated along the line-of-sight, making sphericity a particularly poor assumption in this case, even though the X-ray isophotes are quite round.  Further, \cite{andersson04} find evidence of either large bulk motions of the intracluster gas or of an infalling subcluster, either of which makes it very unlikely the cluster is relaxed and in a state of hydrostatic equilibrium.  Because there are no X-ray halo models that account for this observed substructure, elongation, and dis-equilibrium, we do not at this time include an X-ray constraint.  Though \cite{lemze08} have previously attempted to combine lensing and X-ray observations of Abell 1689, their analysis discards the X-ray temperature profile because it is irregular, further suggesting an unrelaxed state of the cluster, and neglects any non-spherical structure of the cluster.  Thus, at this time it seems a consistent and meaningful combination of X-ray and lensing constraints is not yet possible.

In addition, \cite{lokas06} compiled a spectroscopic study of the cluster from existing observations, but it did not include enough position-velocity pairs to constrain more than the simplest models.  It did however further confirm the complex state of Abell 1689, showing evidence for at least one associated structure in addition to the main halo (it should be noted, however, that \cite{lemze08b} very recently carried out a larger spectroscopic study of the cluster, in which they also identify an infalling subclump but argue that otherwise the cluster appears uniform with little substructure).  While one subclump could perhaps be accounted for by the inclusion of a second clump in the weak lensing mass model in addition to the primary triaxial halo, we found the error contours grow so large when a second spherical halo is added as to be physically insignificant. \cite{limousin07} present strong evidence of a second clump in their strong lensing analysis of the cluster and tightly constrain its position (in the same region where the X-ray data also suggests a major substructure); we attempt a two-component fit with a spherical clump at the location of their Clump 2 in addition to the primary triaxial halo with a $1'$ positional uncertainty; we find that even with the position of the subclump so tightly constrained the weak lensing data has great difficulty constraining the model parameters.  After running the MCMC chains for over four times as long as for the single triaxial halo, convergence is achieved according to the standard statistics, though the likelihood distributions remain very un-smooth, indicating poor constraints.  With that caveat, the marginalized mean mass of the main halo is reduced to $0.92\times 10^{15}$ M$_{\odot}$ while the secondary clump acquires mass $0.29\times 10^{15}$ M$_{\odot}$, with respective concentrations of $C=21.5\pm19.4$ and $C=12.8\pm12.6$.  The errors on the main halo concentration increase by a factor of 2 from the Flat triaxial case, and the $68\%$ confidence contours for the secondary clump include $M=0$.  Thus, the weak lensing data alone cannot meaningfully constrain a subclump so close to the cluster centre, even when its position is externally constrained.

The inclusion of a second clump is not crucial for the weak lensing model, as the secondary peak is located about $1.4'$ from the cluster centre (\cite{limousin07}), inside the central region excluded from the weak lensing analysis.  Though weak lensing is a non-localized phenomenon and thus would not be entirely unaffected by such a clump, the effects would be small compared to those in strong lensing or X-ray analyses that focus on the very centre of the cluster.  Additionally, if the second clump is indicative of a merger along the line-of-sight, which the dynamical and X-ray data suggest it may be, we echo the argument already put forward in \cite{oguri08} that a very prolate triaxial halo is a good zeroth order approximation of such a system.  Indeed, the reality must be somewhere between the extremes of two separate clumps and a fully merged cluster, so a model treating the cluster as single triaxial merged halo is as good or better than any other model currently available to weak lensing analysis, until the dynamical structure of Abell 1689 is better constrained.

In addition to having at least one associated structure, Abell 1689 is likely part of a large-scale matter filament of the cosmic web oriented along the line-of-sight.  This would add significant substructure to the system that cannot be accounted for entirely by a single triaxial halo or the addition of a single associated structure, and may contribute to the apparently high concentration of the system.  We discuss large-scale line-of-sight structure further in \ref{sec:system}.

\begin{figure*}
\centering
\rotatebox{0}{
\resizebox{16cm}{!}
{\includegraphics{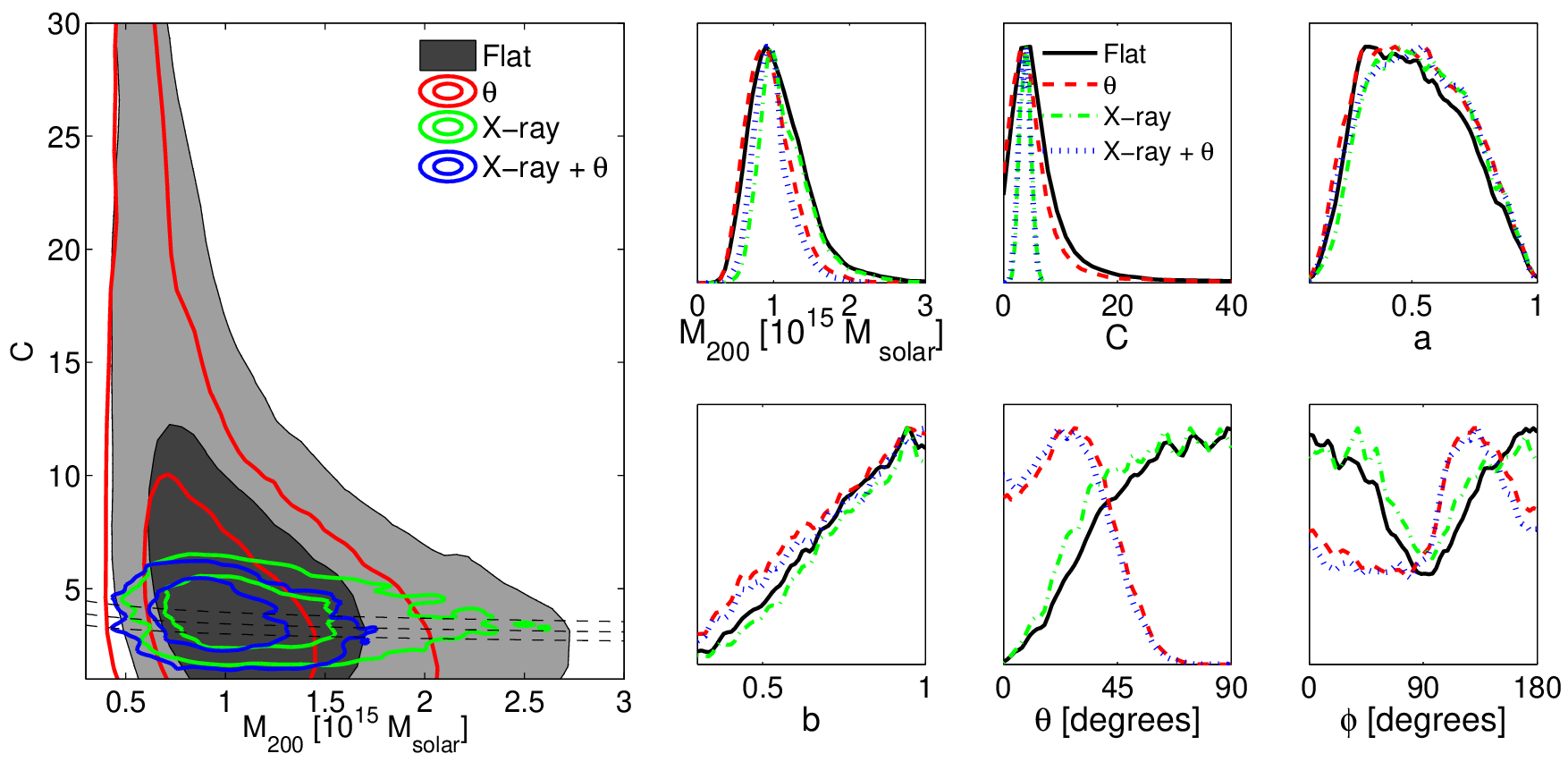}}
}
 \caption{The left-hand panel plots the 2D marginalized 68 and 95 \% confidence contours in the $C-M_{200}$ plane for A1835 under the Flat, $\theta$, and X-ray priors, described in the text.  The dotted lines show the region of parameter space consistent at $1\sigma$ with the mass-concentration relation predicted in $\Lambda$CDM.  The right-hand panels give the corresponding 1D marginalized probability distributions for all six triaxial parameter under each prior.}
 \label{fig:plot8}
\end{figure*}

\subsubsection{Abell 1835}
Abell 1835 has four identified strong lensing features, but these are not spectroscopically confirmed (\cite{schmidt01}; \cite{smith05}).  These suggest an Einstein radius of $31.3''$ for a source redshift of anywhere from $z=0.6$ to $z=3$.  With such poor constraints on the source redshift, application of a direct Einstein radius prior is not possible.  However, because the orientation bias is relatively robust to source redshift -- though it does depend on source redshift, the difference in the percentage of halos with $|\cos\theta| > 0.9$ for a given Einstein radius is only $\sim 5\%$, smaller than the uncertainties in the bias from uncertainties in the cosmological parameters -- we apply the $\theta$ prior as a valid constraint derived from the strong lensing features of the cluster.  For the Einstein radius of Abell 1835, approximately $60\%$ of the halos should have $|\cos\theta| > 0.9$: we implement this as for Abell 1689 via Eq.\ref{eq:tprior}, taking $\sigma_\theta = 0.22$ to reflect the weaker orientation bias for Abell 1835's smaller Einstein radius.

The 2D $M_{200}-C$ confidence contours and 1D marginalized parameter distributions under the $\theta$ prior are shown in Figure \ref{fig:plot8}; the resulting marginalized and maximum-likelihood parameter estimates are given in Tables \ref{tab:table2} and \ref{tab:table3}.  Again, as for Abell 1689, the contours shift to favour lower masses and concentrations, and smaller values of $b$ are favoured while $\phi$ is more tightly constrained.

The core of Abell 1835 has been observed in the X-ray with Chandra (\cite{schmidt01}) and XMM (\cite{majerowicz02}).  The orientation bias is weaker for this halo because of its smaller Einstein radius, and our weak-lensing-only triaxial fits indicate no particular preference toward extreme axis ratios, making the assumption of spherical symmetry of the gravitational potential in the X-ray analysis less problematic.  Moreover, \cite{smith05} argue that the gas properties of the cluster indicate it is relaxed, and there is no significant substructure apparent in lensing or X-ray analyses.  Therefore, the X-ray mass and concentration can reasonably be employed as valuable constraints on the triaxial halo models.  To do so, we convert the best-fitting NFW parameters found in \cite{schmidt01} for an Einstein-De Sitter cosmology to our $\Lambda$CDM cosmology, such that their best-fit parameters $\{r_{200} = 1.28^{+0.45}_{-0.32}$ $h^{-1}$Mpc; $C =  4.0^{+0.54}_{-0.64}\}$ become $\{r_{200} = 1.39^{+0.49}_{-0.34}$ $h^{-1}$Mpc; $C =  3.9^{+0.53}_{-0.63}\}$.  From these best-fitting spherical parameters we calculate the X-ray determined virial mass of the cluster to be $M_{200} = 1.13^{+1.2}_{-0.8}\times10^{15}$ M$_{\odot}$.  (Though there was some discrepancy in the temperature profiles derived from Chandra and XMM due to a potentially unaccounted for solar flare in the Chandra observation (\cite{markevitch02}), the virial mass derived from the XMM analysis by \cite{majerowicz02} is fully consistent with the Chandra mass at $1\sigma$, so we employ only the NFW fit from Chandra rather than the unparameterized constraint from the XMM analysis).  We apply these constraints on the cluster mass and concentration (not on $R_{200}$, since the triaxial and spherical definitions of radius differ) to the triaxial fit to the weak lensing data.  We assign the error distributions Gaussian forms, with errors twice those given for the parameters in order to allow for small residual systematic errors due to the assumptions of spherical symmetry and hydrostatic equilibrium.

Figure \ref{fig:plot8} shows the $M_{200}-C$ confidence contours and 1D marginalized parameter distributions under the X-ray prior; the resulting marginalized and maximum-likelihood parameter estimates are given in Tables \ref{tab:table2} and \ref{tab:table3}. Even with doubled error bars the X-ray signficantly tightens the constraints on both the mass and the concentration.  It very slightly shifts both $a$ and $b$ to higher values, suggesting the spherical assumption of the X-ray model is having some impact on the combined parameter estimates, but that it is small.  Similarly, a shift towards smaller values of $\theta$ implies a slight preference for line-of-sight orientations in which the halo is more likely to look circular on the sky.

The X-ray and weak lensing data agree very well.  In particular, X-ray analyses, with their focus on the central regions of the cluster, typically do very well at accurately constraining the concentration but have more problems accurately constraining the mass (see e.g. \cite{limousin07}); thus the fact that it is the triaxial concentration that is most affected by the application of the X-ray prior, while the triaxial mass stays much the same, further indicates that the X-ray and lensing data are in good agreement, and that the application of the X-ray prior is reasonable and improves the final parameter estimates.

Figure \ref{fig:plot8} also plots the $M_{200}-C$ confidence contours and 1D marginalized parameter distributions under a combined $\theta$ and X-ray prior.  The mass and concentration are both very well constrained, as the preference for line-of-sight orientations slightly lowers the mass and concentration estimates from those under the X-ray prior alone.

\subsubsection{Abell 2204}
Abell 2204 presents no known strong lensing features, and so neither the $\theta$ or $\theta_E$ priors may be applied.  It has been observed in the X-ray with Chandra (\cite{reiprich02}; \cite{sanders05}), XMM (\cite{zhang08}, and Suzaku (\cite{reiprich08}) where it is seen to exhibit complex morphology potentially indicative of a recent merger.  The complex shape of Abell 2204 is further evidenced by our weak-lensing only fits of the triaxial model, which under almost all priors favour highly triaxial halos with extreme axis ratios.  Abell 2204 is thus unrelaxed and of complex shape, making assumptions of sphericity and equilibrium highly suspect; we therefore forego combination of the weak lensing constraints with X-ray observations.

\section{Discussion}
We first compare our fits under the Spherical prior to those of previous work on the three clusters -- that used weak lensing, strong lensing, and X-ray data -- to ascertain the agreement of our lensing catalogue and analysis with other methods and observations.  Once the accuracy of our method and catalogues is so established, we then move on to discuss the impact of triaxiality on the measure of the mass and concentration of A1689, A1835, and A2204.
\begin{figure*}
\centering
\rotatebox{0}{
\resizebox{16cm}{!}
{\includegraphics{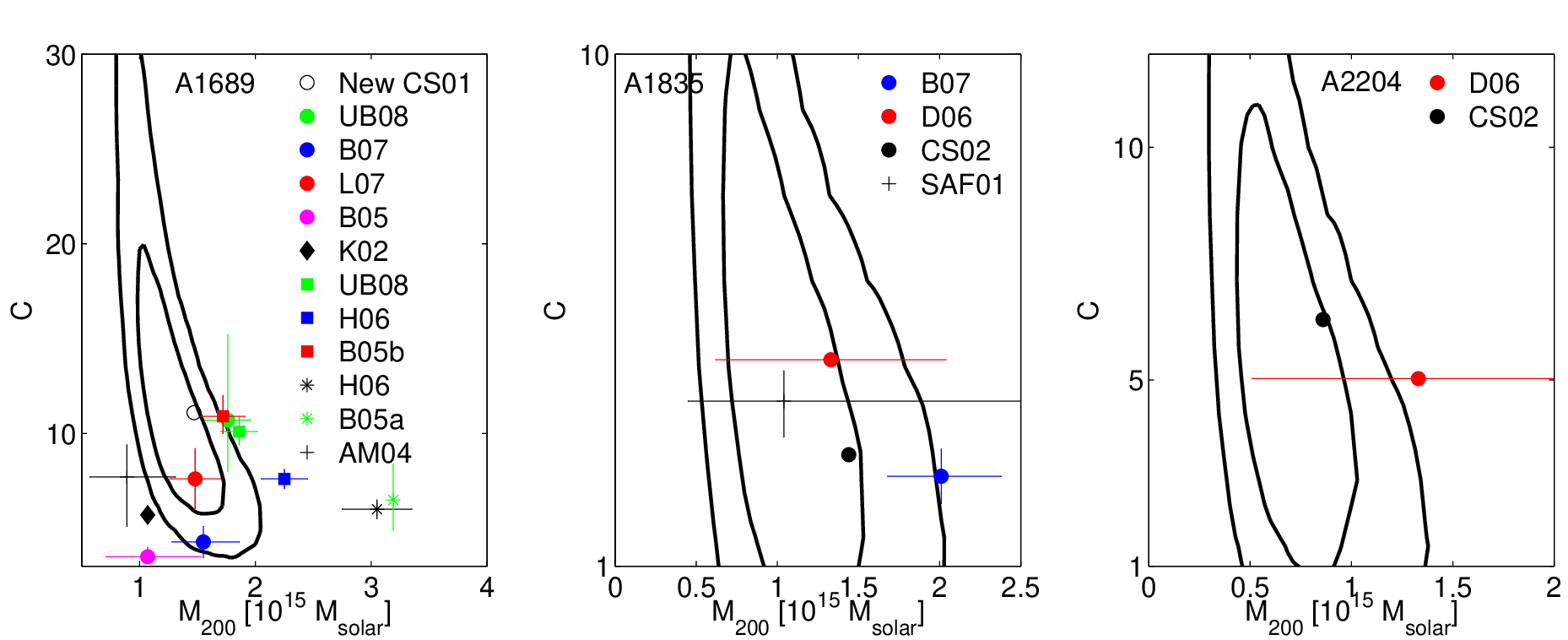}}
}
 \caption{The black contours give the 68 and 95 percent confidence contours for fits under a Spherical prior to the three clusters from this work, compared with the results of other authors fitting spherical models. Circles plot weak lensing-only results, diamonds weak lensing IR results, squares combined weak \& strong lensing analyses, stars strong lensing-only measurements, and crosses X-ray results. The results and authors are compiled in Table \ref{tab:table4}.  The Spherical contours do not represent the best estimates from this work, but are given for more direct comparison with past results to establish the consistency of the MCMC method with previous work.}
 \label{fig:plot9}
\end{figure*}

\subsection{Previous work}
Table \ref{tab:table4} collects the NFW $M_{200}$ and C values published in past work on the three clusters, all converted to a concordance cosmology with $h=0.7$; \cite{limousin07} and \cite{umetsu08} have previously summarised the existing work on Abell 1689 in a similar way.  Figure \ref{fig:plot9} plots most of those results on top of the confidence contours for the fits under a Spherical prior in this work.  Though the Spherical prior fits are NOT our best estimates of the cluster parameters, they are used here in order to more directly compare our work to existing work in order to establish the consistency of the method.  Once this comparison establishes our results as consistent and reliable, we will move on to discuss the fully triaxial results.

\subsubsection{Abell 1689}
The concentration of Abell 1689 has been a source of significant controversy; \cite{limousin07} offers an excellent and thorough discussion of the systematic issues that may affect concentration estimates in the case of this remarkable cluster.  Crucially, contamination of lensing catalogues with very faint cluster members is expected to lead to underestimates of the concentration when magnitude cuts alone are used to distinguish background from foreground and cluster galaxies.  Such was the method used in the early weak lensing studies of Abell 1689 in CS01 and \cite{king02a}.  A version of that same catalogue is employed here, similar to that used in C03, improved with two bands of additional colour data and small improvements to the KSB PSF corrections.  The mean lensing redshift of the background galaxy population is changed in this analysis from $<z_s>=1$ to $<z_s>=0.8$, to account for the more stringent colour cuts in the galaxy population and make use of the improved photo-z information now available from the COSMOS field.  Applying the same fitting technique as in CS01 and C03 to the updated catalogue, we find maximum-likelihood parameter values of $M_{200} = 1.47 \times 10^{15}$ M$_{\odot}$ and $C=11.1$.  The higher mass value compared to the C03 fit can be attributed primarily to the change in the mean background source redshift assumed in the analysis: overestimating the mean background redshift as was done in C03 shifts the $M_{200}-C$ contours to lower masses.  In our MCMC maximum-likelihood fit under a spherical prior, we obtain marginalized mean parameter values $M_{200} = 1.27\pm0.24\times 10^{15}$ M$_{\odot}$ and $C=13.3\pm7.6$, and approximate maximum-likelihood parameters $M_{200} = 1.33^{+0.44}_{-0.34} \times 10^{15}$ M$_{\odot}$ and  $C=10.0^{+9.3}_{-4.0}$ which are consistent with this most recent result at $1\sigma$.  The concentration values have increased with every new iteration of the catalogue, as expected as the level of contamination by faint cluster members decreases.

\cite{king02b} performed a weak lensing analysis employing infrared data.  While the infrared provides an order of magnitude fewer background galaxies, it allows for a more secure determination of the cluster red sequence and removal of foreground galaxies.  The mass and concentration yielded in that study are both lower than those we find here; however, that is expected because that analysis did not include the $\sim 1.1$ shear correction of \cite{bacon01}. 

Our marginalized spherical mass agrees well with those of the weak lensing results of \cite{bardeau05} and \cite{limousin07}, both using CFHT observations of the cluster, and is generally lower than those using the Subaru observation.  However, our mean concentration is generally higher than those inferred from the CFHT observations, and more consistent with those of the combined weak and strong analyses of \cite{umetsu08} and \cite{broadhurst05b}, though not so high as the recent weak-lensing only \cite{medezinski07} result, which all employ the Subaru data.  Overall, our results fit well into the body of weak lensing and combined strong \& weak lensing work that exists on the cluster.  Compared to the strong lensing-only results, our spherical mass is low compared to those found using ACS data, as with all of the weak lensing results.

Compared to the X-ray result, our mass and concentration are both high, though marginally consistent at $1\sigma$.  This is unsurprising as \cite{andersson04} find evidence for an ongoing or recent merger near the cluster centre, a finding supported by strong lensing and flexion (\cite{leonard07}) measurements as well, and argue that X-ray analyses will underestimate the total mass in such cases.

\subsubsection{Abell 1835}
The catalogue employed for Abell 1835 is improved from that published in CS02 and C03, as the addition of two colour bands from CFHT observations allowed for a much more reliable discrimination between cluster, foreground, and background galaxies.  Using the catalogue constructed using a magnitude cut CS02 reported NFW parameter values $M_{200} = 1.44 \times 10^{15}$ M$_{\odot}$ and $C=2.96$.  Using the new catalogue, and a lower mean source redshift $<z_s>=0.8$, compared to the value of $<z_s>=1.0$ used in the CS02 and C03 analyses, we find for a spherical NFW marginalized mean parameters $M_{200} =1.01\pm0.30\times 10^{15}$ M$_{\odot}$ and $C=6.8^{+9.5}_{-6.8}$ and approximate maximum-likelihood parameters $M_{200} =1.07^{+0.55}_{-0.39}\times 10^{15}$ M$_{\odot}$ and $C = 3.6^{+5.4}_{-2.0}$.  Our recovered concentration is higher, as expected given the significant improvement in the removal of faint galaxy contamination with the addition of two colour bands.  The recovered mass, however, is lower, though still comfortably consistent with our new results at $1\sigma$.

Looking to other weak lensing studies, our spherical mass is lower than that of \cite{bardeau07}, and our concentration higher.  However, our mass and concentration are in agreement with the \cite{dahle06} result at $1\sigma$.  Further, \cite{smith05} find a total projected mass within 500 kpc of the cluster centre of $M=(2.9\pm0.6) \times 10^{14}$ $h^{-1}$M$_{\odot}$, for an Einstein-De Sitter cosmology (they argue converting to a concordance cosmology would induce a change of no more than $10\%$, less than the error bars).  For the same angular radius, our model in the concordance cosmology gives a projected mass of $M=3.0\times 10^{14}$ $h^{-1}$M$_{\odot}$ under a Flat prior, and for more direct comparison, $M=2.7\times 10^{14}$ $h^{-1}$M$_{\odot}$ under a Spherical prior, both in good agreement with the Smith result.  

Our spherical mass and concentration estimates also agree very well with the parameter estimates derived from Chandra observations under assumptions of spherical symmetry (\cite{schmidt01}). \cite{zhang08} studied Abell 1835 with XMM and found a mass of $M_{500} = (5.90\pm1.72) \times 10^{14}$ $h^{-1}$M$_{\odot}$ (where $M_{500}$ is the mass contained within a sphere with mean density 500 times the critical density); our marginalized mean model under a Flat prior gives $M_{500} = 5.8 \times 10^{14}$ $h^{-1}$M$_{\odot}$, also in good agreement.
\begin{table*}
\centering
\caption{Mass and Concentration estimates from previous work.  All lensing- and X-ray- derived values assumed spherical symmetry, and all X-ray analyses also assumed hydrostatic equilibrium.  Values in parentheses were fixed during fitting.}
\label{tab:table4}
\begin{tabular}[t!]{ccccl}
\hline
$M_{200}$ $[10^{15}$ M$_{\odot}]$&$C$&Method&Instrument&Author\\
\hline
A1689&&&&\\
1.47&11.10&WL&ESO/MPG WFI&CS01 method with current catalogue\\
$1.76\pm0.20$&$10.7^{+4.5+}_{-2.7}$&WL&Subaru&\cite{umetsu08}\\
--&$22.1^{+2.9}_{-4.7}$&WL&Subaru&\cite{medezinski07}\\
$1.55^{+0.31}_{-0.27}$&$4.28\pm0.82$&WL&CFHT&\cite{bardeau07}\\
$1.48\pm0.22$&$7.6\pm1.6$&WL&CFHT&\cite{limousin07}\\
$1.07^{+0.46}_{-0.36}$&$3.5^{+0.5}_{-0.3}$&WL&CFHT&\cite{bardeau05}\\
1.13&7.9&WL&ESO/MPG WFI&C03\\
1.07&5.7&WL (Infrared)&NTT SOFI&\cite{king02b}\\
0.85&4.8&WL&ESO/MPG WFI&\cite{king02a}\\
$1.86\pm0.16$&$10.1^{+0.8}_{-0.7}$&WL + SL&Subaru + ACS&\cite{umetsu08}\\
$2.25\pm0.20$&$7.6\pm0.5$&WL + SL&Subaru + ACS&\cite{halkola06}\\
$1.72\pm0.19$&$10.9^{+1.1}_{-0.9}$&WL + SL&Subaru + ACS&\cite{broadhurst05b}\\
--&$6.0\pm0.6$ ($3\sigma$)&SL&ACS&\cite{limousin07}\\
$3.05\pm0.30$&$6.0\pm0.5$&SL&ACS&\cite{halkola06}\\
3.19&$6.5^{+1.9}_{-1.6}$&SL&ACS&\cite{broadhurst05a}\\
$0.89^{+0.42}_{-0.32}$&$7.7^{+1.7}_{-2.6}$&X-ray&XMM&\cite{andersson04}\\
\hline
A1835&&&&\\
$2.01^{+0.37}_{-0.33}$&$2.58\pm0.48$&WL&CFHT&\cite{bardeau07}\\
$1.33\pm0.71$&(4.63)&WL&NOD/UHT&\cite{dahle06}\\
1.44&2.96&WL&ESO/MPG WFI&CS02\\
0.83&--&X-ray&XMM&\cite{majerowicz02}\\
$1.04^{+1.52}_{-0.59}$&$3.90^{+0.53}_{-0.63}$&X-ray&Chandra&\cite{schmidt01}\\
\hline
A2204&&&&\\
$1.33\pm0.82$&(5.03)&WL&NOD/UHT&\cite{dahle06}\\
0.86&6.3&WL&ESO/MPG WFI&CS02\\
\hline
\end{tabular}
\end{table*} 

\subsubsection{Abell 2204}
Though the colour selection in the catalogue used in this work is the same as that in CS02 and C03, as with the other two catalogues the PSF correction is improved and a lower mean background source redshift is employed to better reflect the expected mean redshift of the population after the colour selection.  Our marginalized mean parameters ($M_{200} =0.69\pm0.21\times 10^{15}$ M$_{\odot}$; $C=6.6^{+6.7}_{-6.6}$) and maximum-likelihood parameters ($M_{200} =0.71^{+0.38}_{-0.26}\times 10^{15}$ M$_{\odot}$; $C=4.5^{+5.4}_{-2.4}$) under a Spherical prior agree with those of CS02 at $1\sigma$.  Our mass is lower than that reported by \cite{dahle06}, but within their (large) $1\sigma$ errors.  

\cite{zhang08} studied Abell 2204 with XMM and found an X-ray mass $M_{500} = (4.12\pm1.12) \times 10^{14}$ $h^{-1}$M$_{\odot}$; our marginalized model under a Flat prior gives $M_{500} = 3.7 \times 10^{14}$ $h^{-1}$M$_{\odot}$, fully consistent with the X-ray result.  

Thus, though there continues to be a significant scatter in the parameter values returned for all three clusters, our parameter estimates under a spherical prior or in projection (where 3D structure does not matter) are all comfortably situated within the errors of previous work utilizing weak lensing, strong lensing, and X-ray observations.  The improvements in background galaxy discrimination are apparent in the increased concentrations of Abell 1689 and Abell 1835.  We are therefore confident our catalogues and MCMC sampler are returning reasonably accurate estimates of the mass profiles of these clusters, and understand how those profiles compare to those derived in previous work.  Now we are equipped to examine the mass profiles of each of the clusters freed from the unphysical assumption of spherical symmetry.

\subsection{Triaxial parameter estimates}
The parameter estimates for halo mass and concentration using the triaxial NFW model should provide a better estimate of the true parameters and errors of lensing galaxy clusters, as the model more closely reflects our true beliefs regarding the shape and structure of clusters.  Generally, we find that triaxial models under the most general Flat prior return higher mass and concentration estimates than their spherical counterparts, but that otherwise the impact of triaxial fitting under other more stringent priors, whether drawn from theory or independent observational constraints, can shift parameters estimates in either direction from the typical spherical values.  

The question of which prior is most appropriate for a given question remains open, and crucially important for this underconstrained problem.  Generally, in the case where only weak lensing information is available, the most general Flat prior seems most useful, as it makes minimal assumptions about the halo shapes and parameter relations.  Until the question of overconcentration is resolved, it seems wise to avoid any prior that ties the concentration to the mass, as this may hide real discrepancies.  Given that, the application even of a less controversial Mass prior is unwise, because the shape of the mass-concentration degeneracy automatically requires higher concentration values as mass values are pushed to their lower limits.  This artificially exacerbates the mass-concentration relation discrepancy.  A mixed $MC$ prior may be of great use in the future if the relation is better understood and constrained, as it can provide simple physical constraints on this underconstrained problem without requiring independent observational data.  However, it is only valid once it represents our true prior expectation for galaxy cluster structure, a status which current observational data does not yet confer.

Similarly, employing a spherical prior imposes a prior constraint that does not represent our true best predictions for cluster structures.  The $M_{200}-C$ contours show that generally the contours under the Spherical prior are enclosed by those of the Flat triaxial models, though they occupy different subregions of the triaxial parameter space depending on the structure of the lensing cluster.  Thus, they not only give errors that are too small and unrepresentative of the true uncertainties in the problem, but unpredictably bias the mass and concentration estimates either high or low.  Though these biases were shown to cancel with large-scale averaging in CK08, they are a large problem when studying individual clusters.  Similarly, the overly-optimistic error estimates make proper comparisons of parameter values between different works inaccurate, as fully consistent results may appear excluded by inaccurately tight error bars; this is even more important because fitting spherical models to non-spherical data makes the details of the fitting method and its order of operations in measuring, averaging, and weighting shear far more important and likely to affect the final parameter estimate.  The inaccurate description of errors is also important when attempting to characterize the scatter in mass-observable relations, crucial knowledge for constraining the cluster mass function and with it cosmological questions such as the normalization of the matter power spectrum and the nature of dark energy.
\begin{figure}
\centering
\rotatebox{0}{
\resizebox{8cm}{!}
{\includegraphics{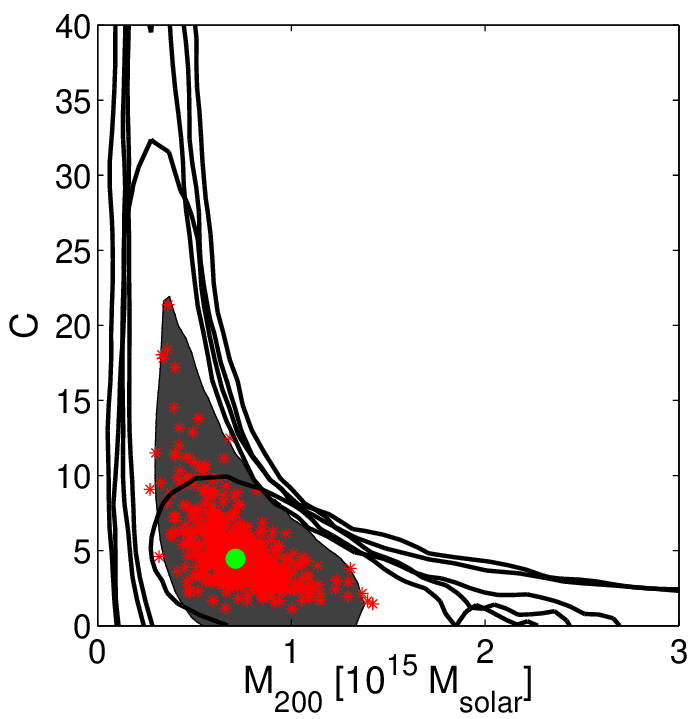}}
}
 \caption{The red stars plot the maximum-likelihood $M_{200}$ and $C$ parameter values obtained fitting a spherical NFW to 500 randomly chosen halves of the Abell 2204 catalogue.  The green circle gives the maximum-likelihood parameter values, and the shaded black region gives the $2\sigma$ confidence region, obtained fitting the same spherical NFW to the whole catalogue.  The black contours give the $2\sigma$ confidence contours for 5 of the 500 half-catalogue realizations; the mass sheet degeneracy becomes very strong when the signal is weak, and causes the posterior probability distribution to develop very long tails to high concentrations and masses.}
 \label{fig:plot10}
\end{figure}

As \cite{oguri08} convincingly demonstrated over a large population of triaxial halos, and as we argued in CK08, the appearance of circular symmetry in a projected cluster DOES NOT justify the assumption of sphericity, most especially in strong lensing clusters.  \cite{oguri08} found that the strongest lensing clusters in the universe are expected to be more triaxial, but look more circular in projection, than their weaker counterparts, as line-of-sight projection boosts the lensing strength; CK08 showed that such halos leave no signature in their lensing signal to differentiate them from truly spherical halos, making it impossible to determine beforehand the degree of triaxiality of a given system.  Thus, for all systems, and especially powerful strong lens systems such as Abell 1689, assuming spherical symmetry is never justified theoretically.

When external information is available, such as strong lensing, X-ray, or dynamical data, this MCMC method allows for the straightforward combination of those external constraints with the weak lensing data.  However, as emphasized in Sec. \ref{sec:extpriors}, care must be taken that those external constraints do not reintroduce the very assumptions the triaxial model intends to avoid.  In the case of strong lensing, this is most easily done through the $\theta$ orientation prior, which comes from a robust theoretical prediction favouring line-of-sight oriented halos to have the highest cluster lensing signals in the universe.  The Einstein radius prior is in some ways preferable in that it requires no prediction from CDM theory, but is problematic in that, applied to a dark matter only profile such as the NFW, it neglects the impact of baryons in the very centre of the cluster that are expected to significantly boost the strong lensing signal. This leads to a strong upward shift in the concentration value, which is unlikely to accurately represent the structure of the dark matter component of the cluster.  X-ray constraints may be applied when a halo appears to have a low degree of triaxiality, making the assumption of spherical symmetry of the (always more spherical) 3D gravitational potential less problematic, and when the cluster appears relaxed and undisturbed.

More detailed constraints may be applied by their direct inclusion in the likelihood function constraining the triaxial model.  For example, strong lensing may be used to directly constrain the mass and concentration if the strongly lensed features, or the 2D convergence map constrained by them (constructed with no assumptions about the 3D structure of the halo), are used to directly constrain the 3D triaxial model.  Doing so requires careful calibration of the weighting between the weak and strong lensing data, an optimization that has not yet been fully examined but has been begun by various groups (\cite{bradac05}; \cite{bradac06}; \cite{diego07}), and should be a focus of the next generation of galaxy cluster studies.

\subsection{Remaining systematics}\label{sec:system}
Though the parameters and the errors presented in this work now include the crucial contribution of halo triaxiality, there remain several outstanding systematics.

One of these is uncertainty in the mean redshift of the background galaxy population, changes in which would lead to changes in the mass estimates of several percent.  Another is residual contamination by cluster galaxies; though the colour cuts made in all three catalogues are very stringent, there is still the possibility of contamination by a few cluster members.  As with the uncertainties in the mean redshift of the background population, the effects should be negligible.

More importantly, stringent colour cuts to remove contamination often result in a significant decrease in the number density of background sources (for example, the catalogue from Abell 1689 used in this work had a source density of only 7/arcmin$^2$).  Low number density affects maximum-likelihood values and marginalized parameters significantly because of the large and non-Gaussian mass-sheet degeneracy between the NFW mass and concentration: mass and concentration constraints from low S/N data are poor because the tails of the highly asymmetric $M_{200}-C$ posterior probability distribution elongate dramatically.  Figure \ref{fig:plot10} shows this effect, taking 500 randomly chosen halves of the Abell 2204 catalogue and treating them each as individual realizations to which a spherical NFW is fit.  Red stars plot the distribution of recovered maximum-likelihood parameters, compared to the $2\sigma$ confidence contour for the spherical model fit to the complete catalogue.  The mean marginalized parameter values are even worse affected, with recovered values of concentration up to six times greater than the mean from the complete catalogue, because they are very sensitive to the long tails of the posterior when the signal is very weak, illustrated for five realizations in the Figure.  The mean marginalized parameters would be expected to be {\it more} robust than their maximum-likelihood counterparts were the errors Gaussian; their sensitivity to the random removal of half the lensing signal reflects the very non-Gaussian errors induced by the mass-sheet degeneracy.  Because marginalized parameters are more meaningful in underconstrained problems such as that of fitting 3D mass models to lensing data, this is an important issue.  More efficient ways of removing foreground contamination are therefore very valuable, such as the Bayesian method of \cite{limousin07}, or that of \cite{medezinski07} that increases the efficiency of background selection by studying the radial tangential shear profiles of various colour populations, or improved methods of colour discrimination based on detailed photometric studies of the background galaxy population.  Future lensing work, including studies using the MCMC method employed here, can only be improved by increasing the background number density reliably available for inclusion in the lensing analysis.

Our triaxial model includes no substructure, while we know galaxy cluster scale halos often contain subclumps, and certainly contain mass concentrations at the positions of the cluster galaxies and their accompanying dark matter halos.  The weak lensing data does not have the resolution or signal-to-noise to constrain these; for example, \cite{marshall06} found the Bayesian evidence supported fitting usually at most two components to cluster-scale substructures.   In the future, in combination with strong lensing or flexion, where detailed models including substructure are frequently constructed (e.g. \cite{limousin07}; \cite{jullo07}; \cite{leonard07}), it should be possible to constrain more of the substructure of clusters.  However, \cite{clowe04} showed that the effects of substructure on parameter estimate are small compared to those of triaxiality, and so we expect the inclusion of multiple components to be a minor perturbation to the results we present in this paper.

Finally, our quoted errors do not include a contribution from uncorrelated line-of-sight structure, shown by \cite{hoekstra03} to induce an unbiased scatter in parameter estimates that can increase the errors on concentration and mass by up to a factor of 2.  \cite{dodelson04} demonstrated that this can be somewhat mitigated by including a noise term for large scale structure in the lensing analysis; building on the greater coherence scale of large scale structure noise compared with the noise associated with the intrinsic ellipticity dispersion of galaxies, Dodelson notes that the errors on cluster mass can be reduced by $\sim 50\%$ for wide-field data, though this has yet to be implemented in any weak lensing analysis.

\subsection{Bayesian Evidence}
We choose not to use the Bayesian evidence to discriminate between our fits under different priors, because we are not attempting to choose between them as models of the universe.  The limitations of the evidence are clear when the Flat vs Spherical cases are considered: in many cases the evidence would likely favour a Spherical over a Flat prior, due to a significant decrease in available parameter space coupled with a relatively small decrease in likelihood values under the Spherical prior, as seen in the very similar reduced $\chi^2$ values between all priors. However, unlike in cases of fitting multiple halos to account for substructure, or in another context, adding additional parameters to cosmological models, we know a priori from physical observations that a triaxial model is a better model than a spherical model -- we see clearly in non-parametric lensing mass maps that galaxy clusters have significant levels of ellipticity and are not spherical!  Thus, the prior on the spherical {\it model} is close to zero, but difficult to quantify.  

The various other theoretical priors studied in this paper are all derived from a single CDM model, and so treating them as separate models to choose between is inappropriate.  We study them here independently to understand how various aspects of the CDM model interact with the lensing data; in application to a large survey, all priors that represent the true prior expectation for the problem without interfering with the scientific questions asked by the study should be employed simultaneously.  Similarly, external constraints from strong lensing or X-ray do not represent different models to be selected between, only appropriate or inappropriate measures to constrain the triaxial models.

The evidence is a powerful tool for model discrimination, but is not the appropriate method for choosing priors on a single physical model, well-founded in theory and observations.

\subsection{Summary}
We fit a fully triaxial NFW model to weak lensing data from massive galaxy clusters Abell 1689, Abell 1835, and Abell 2204, deriving parameter estimates and errors under a range of theoretical and observational priors.  Under a relatively weak prior drawn from the strong lensing orientation bias for Abell 1689, that same strong lensing prior combined with an X-ray derived mass and concentration constraint for Abell 1835, and under the very general Flat prior for Abell 2204, our best mean parameter estimates in a concordance cosmology with $h=0.7$ are
\begin{itemize}
\item Abell 1689: $M_{200}=(1.18 \pm .23)\times10^{15}$ M$_{\odot}$; $C=12.2\pm 6.7$;
\item Abell 1835: $M_{200}=(0.98 \pm .25)\times10^{15}$ M$_{\odot}$; $C=3.7\pm 1.0$;
\item Abell 2204: $M_{200}=(0.72 \pm .27)\times10^{15}$ M$_{\odot}$; $C=7.1\pm 6.2$;
\end{itemize}
All results are consistent with previous work, but importantly with larger error contours that better reflect the true uncertainty in mass profile estimates.

Triaxiality does not easily return Abell 1689 to the fold of average CDM clusters; the predicted CDM mass-concentration relation is enclosed in its $2\sigma$ $M_{200}-C$ confidence contour under most priors, but is excluded under some observational priors, making Abell 1689 at the moment weakly consistent with the predictions of CDM at $2\sigma$.  However, consistency does not mean truth: the primary result of this work is to demonstrate the necessity for improved methods of combining diverse observational constraints to counter the very large uncertainties that accompany fits of triaxial NFW profiles to galaxy cluster lenses.  That way lies the future of galaxy cluster studies, whether of individual clusters or large populations and mass-observable relations: without such optimized and robust combination methods, most results will continue to be consistent with one another and with theory simply through the size of their errors, rather than through physical concordance.

\section*{Acknowledgements}
This work was supported by the National Science Foundation, the Marshall Foundation, and the Cambridge Overseas Trust (VLC) and the Royal Society (LJK).  We thank Jean-Paul Kneib and Oliver Czoske for making the reduced CFHT images of A1689 and A1835 available, and Helen Russell, Hans Boehringer, Graham Smith, and Antony Lewis for very useful discussions.

\appendix

\section{Lensing through triaxial halos}\label{sec:appa}

Following OLS, the triaxial halo is projected onto the plane of the sky to find its projected elliptical isodensity contours as a function of the halo's axis ratios and orientation angles ($\theta$, $\phi$) with respect to the the observer's line-of-sight.\footnote{Although we set $c=1$, we keep $c$ as a variable in our notation for consistency with OLS.}    The elliptical radius is given by
\begin{equation}\zeta^2 = \frac{X^2}{q_X^2} + \frac{Y^2}{q_Y^2}\label{eq:xi}\end{equation}
where $(X,Y)$ are physical coordinates on the sky with respect to the centre of the halo,
\begin{eqnarray}
q_X^2&=&\frac{2f}{\mathcal{A}+\mathcal{C} - \sqrt{(\mathcal{A}-\mathcal{C})^2 + \mathcal{B}^2}}\\
q_Y^2&=&\frac{2f}{\mathcal{A}+\mathcal{C} + \sqrt{(\mathcal{A}-\mathcal{C})^2 + \mathcal{B}^2}} \end{eqnarray}
where
\begin{equation}f = \sin^2\theta\left(\frac{c^2}{a^2}\cos^2\phi + \frac{c^2}{b^2}\sin^2\phi\right) + \cos^2\theta,\label{eq:f}\end{equation}
and
\begin{eqnarray}
\mathcal{A}&=&\cos^2\theta\left(\frac{c^2}{a^2}\sin^2\phi + \frac{c^2}{b^2}\cos^2\phi\right) + \frac{c^2}{a^2}\frac{c^2}{b^2}\sin^2\theta,\\
\mathcal{B}&=&\cos\theta\sin 2\phi\left(\frac{c^2}{a^2} - \frac{c^2}{b^2}\right),\\
\mathcal{C}&=&\frac{c^2}{b^2}\sin^2\phi + \frac{c^2}{a^2}\cos^2\phi.
\end{eqnarray}
The axis ratio $q$ of the elliptical contours is then given by
\begin{equation}q = \frac{q_Y}{q_X}\label{eq:q}\end{equation}
and their orientation angle $\Psi$ on the sky by
\begin{equation}\Psi = \frac{1}{2}\tan ^{-1}\frac{\mathcal{B}}{\mathcal{A}-\mathcal{C}}~~~(q_X \ge q_Y).\label{eq:psi}\end{equation}

Here we diverge slightly from OLS's treatment as we are interested not in deflection angles but in the lensing shear and convergence, both combinations of second derivatives of the lensing potential $\Phi$ (commas indicate differentiation):
\begin{eqnarray}
\gamma_1&=& \frac{1}{2}\left(\Phi_{,XX} - \Phi_{,YY}\right),\\
\gamma_2 &=&\Phi_{,XY},\\
\kappa &= &\frac{1}{2}\left(\Phi_{,XX} + \Phi_{,YY}\right).
\label{eq:gammakappa}\end{eqnarray}
These derivatives are calculated as functions of integrals of the spherical convergence $\kappa(\zeta)$ (see e.g. \cite{bartelmann96} for a full treatment of weak lensing by a spherical NFW profile) following the method of \cite{schramm90} and \cite{keeton01}, normalised by a factor of $1/\sqrt{f}$ from Equation~\ref{eq:f} (see OLS for the derivation of this normalisation)
\begin{eqnarray}
\Phi_{,XX} &= &2qX^2K_0 + qJ_0,\\
\Phi_{,YY} &= &2qY^2K_2 + qJ_1,\\
\Phi_{,XY} &= &2qXYK_1,\end{eqnarray}
where
\begin{eqnarray}K_n(X,Y)& =&\frac{1}{\sqrt{f}}\int_0^1 \frac{u\kappa'(\zeta(u)^2)}{[1 - (1-q^2)u]^{n+1/2}}du,\label{eq:K}\\
J_n(X,Y)& =&\frac{1}{\sqrt{f}}\int_0^1 \frac{\kappa(\zeta(u)^2)}{[1 - (1-q^2)u]^{n+1/2}}du,
\end{eqnarray}
and
\begin{equation}\zeta(u)^2 = \frac{u}{q_X}\left(X^2 + \frac{Y^2}{1 - (1-q^2)u}\right).\end{equation}

Note that our radial variable $\zeta$ appears different from Keeton's $\xi$ because it is defined in terms of two axis ratios $q_X$ and $q_Y$ rather than one $q$: $\zeta = \xi/q_X$.  This reflects a dependence on the 3D structure of the cluster; for example, extended structure along the line of sight decreases $q_X$ and thus increases the convergence and shear at a given $(X,Y)$.

\onecolumn
\section{Priors from Simulations: Shaw Prior} \label{sec:appb}

We define the Shaw prior, plotted in Figure \ref{fig:plot1}, by fitting polynomials to the data points of Figure 14 in \cite{shaw06}:

$p(b) = \left\{\begin{array}{ll} 0 & \textrm{if }b < 0.5\\\left[1.6329b^5 - 7.9775b^4+9.3414b^3 - 6.6558b^2+2.2964b - .3088\right]\times 10^3 & \textrm{if }0.5 \leq b \leq 1.0\end{array}\right.$

and

$p\left(\frac{a}{b}\right) = \left\{\begin{array}{ll}0& \textrm{if }\frac{a}{b} < 0.65\\
\left[5.76647\left(\frac{a}{b}\right)^6 - 2.459265\left(\frac{a}{b}\right)^5 + 42.3154\left(\frac{a}{b}\right)^4 - 37.2765\left(\frac{a}{b}\right)^3\right.&\textrm{if }0.65 \leq \frac{a}{b} \leq 1.0.\\
\left.~~~~~+ 17.4650\left(\frac{a}{b}\right)^2 - 4.00238\left(\frac{a}{b}\right)+ .32462\right]\times 10^4&\end{array}\right.$

\bsp

\label{lastpage}

\end{document}